\documentclass[%
  reprint,
  amsmath,amssymb,
  aps,
  prapplied,
  superscriptaddress,
  longbibliography]{revtex4-2}

\usepackage{tikz}
\usepackage{amsmath} 

\usepackage{graphicx}

\usepackage{dcolumn}
\usepackage{bm}
\usepackage[english]{babel}
\usepackage{hyperref}
\usepackage{url}
\usepackage{xcolor}

\DeclareUnicodeCharacter{2009}{\,}

\begin{document}

\preprint{APS/123-QED}

\title{The Role of Exceptional Points and Transmission Peak Degeneracies in Non-Hermitian Sensing}

\author{Alexander S. Carney}
\affiliation{\mbox{Thayer School of Engineering, Dartmouth College, 15 Thayer Drive, Hanover, New Hampshire 03755, USA}}

\author{Juan S. Salcedo-Gallo}
\affiliation{\mbox{Thayer School of Engineering, Dartmouth College, 15 Thayer Drive, Hanover, New Hampshire 03755, USA}}

\author{Salil K. Bedkihal}
\affiliation{\mbox{Thayer School of Engineering, Dartmouth College, 15 Thayer Drive, Hanover, New Hampshire 03755, USA}}

\author{Mattias Fitzpatrick}
\affiliation{\mbox{Thayer School of Engineering, Dartmouth College, 15 Thayer Drive, Hanover, New Hampshire 03755, USA}}
\affiliation{\mbox{Department of Physics and Astronomy, Dartmouth College, 6127 Wilder Laboratory, Hanover, New Hampshire 03755, USA}}

\begin{abstract}
Transmission peak degeneracies (TPDs) have emerged as a promising alternative to exceptional points (EPs) for non-Hermitian sensing, providing square-root frequency splitting without the eigenbasis collapse and associated noise amplification that limit EP sensors. However, existing treatments of TPDs remain fragmented, lacking a unified theoretical framework, systematic figures of merit, or design principles for practical implementation. Here, we develop a comprehensive theory of two-dimensional TPDs that clarifies their relationship to EPs, maps their locations in parameter space, and provides analytic figures of merit for sensor design. We validate our theory using a tunable cavity-magnonics platform with in situ control of mode frequency, dissipation, and complex coupling via an effective synthetic gauge field. Our platform enables systematic exploration of six representative EP–TPD configurations spanning $\mathcal{PT}$-symmetric, anti-$\mathcal{PT}$-symmetric and anyonic-$\mathcal{PT}$-symmetric regimes. Crucially, we show that TPDs, unlike EPs, retain square-root splitting even under nuisance parameter drift through generalized transmission extrema degeneracies (TEDs). We further identify specific robust TPD configurations that minimize the impact of nuisance drift. These findings establish a unified theoretical and experimental framework for TPD-based non-Hermitian sensing.
\end{abstract}

\date{\today}
\maketitle

\section{\label{sec:intro} Introduction}

The past decade has seen sustained theoretical and experimental efforts to leverage exceptional points (EPs) for enhanced sensing \cite{wiersig_review_2020, wiersig_sensors_2016, wiersig_enhancing_2014, park_symmetry-breaking-induced_2020, kononchuk_exceptional-point-based_2022, suntharalingam_noise_2023}. The appeal of EP sensors is sublinear sensing: near an EP of order $n \geq 2$, the eigenvalue splitting scales as $\Delta_\lambda \propto \epsilon^{1/n}$ with perturbation $\epsilon$. Therefore, the corresponding susceptibility $\partial(\Delta_\lambda)/\partial\epsilon \propto 1/{\epsilon^{1 - 1/n}}$ diverges as $\epsilon\to0$, suggesting a route to dramatically enhance the sensor response. EPs have been studied in numerous experimental platforms including $\mathcal{PT}$-symmetric optical resonators \cite{chen_parity-time-symmetric_2018, ozdemir_paritytime_2019, miri_exceptional_2019}, electrical and superconducting circuits \cite{kononchuk_exceptional-point-based_2022, lu_harnessing_2025, dong_sensitive_2019, huerta-morales_generating_2023, naghiloo_quantum_2019, chen_decoherence-induced_2022}, and cavity-magnonic \cite{zhang_observation_2017} systems, as well as anti-$\mathcal{PT}$-symmetric circuits \cite{choi_observation_2018} and anyonic-$\mathcal{PT}$-symmetric lasers \cite{arwas_anyonic-parity-time_2022}. However, translating EP physics into practical sensors has faced considerable hurdles. In practice, EPs are isolated degeneracies in high-dimensional parameter spaces, and any deviation in a nuisance (non-sensing) parameter instantly lifts the degeneracy and removes the square-root eigenvalue splitting, rapidly diminishing the enhanced response. Furthermore, although early demonstrations of EP sensors show superior susceptibility to perturbations \cite{chen_exceptional_2017}, the complete eigenbasis collapse at EPs that yields the square-root splitting response inherently amplifies noise \cite{wang_petermann-factor_2020}. This intrinsic noise amplification at EPs, quantified by the divergence of the Petermann factor \cite{siegman_excess_1989}, fundamentally limits any improvement in the signal-to-noise ratio (SNR) for EP-based sensors \cite{langbein_no_2018, loughlin_exceptional-point_2024}. Consequently, the viability of frequency-splitting-based EP sensors has been the subject of contentious debate \cite{mortensen_fluctuations_2018, wang_petermann-factor_2020, wiersig_prospects_2020, grant_rotation_2021, zhang_quantum_2019, lau_fundamental_2018}.

Recent efforts have aimed to circumvent the fundamental noise–susceptibility tradeoff at EPs \cite{lu_harnessing_2025, kononchuk_exceptional-point-based_2022, xu_single-cavity_2024, geng_discrepancy_2021}. Specifically, a promising and experimentally accessible solution is to use degeneracies in the transmission spectrum rather than the eigenspectrum \cite{geng_discrepancy_2021}. Transmission peak degeneracies (TPDs) reproduce the same square-root splitting of EPs while generally maintaining linearity and a complete eigenbasis, and recent experiments have demonstrated that TPD-based sensors achieve improved SNR performance \cite{lu_harnessing_2025, kononchuk_exceptional-point-based_2022, xu_single-cavity_2024}. However, existing literature has largely treated TPDs as static, isolated phenomena, overlooking the continuous parameter landscape that connects them. This fragmented view prevents a principled comparison of TPD candidates for building sensors. Unlike EPs, whose properties are topologically fixed, TPDs exist within a tunable design space where key figures of merit can be actively engineered through coupling phase and dissipation rates. Furthermore, while demonstrations of TPD sensors have achieved improved SNR, TPDs retain the notorious fragility to nuisance fluctuations characteristic of EPs. Previous studies have either ignored this vulnerability or addressed it with closed-loop feedback \cite{lu_harnessing_2025}, leaving the underlying mechanism unexplored. More broadly, the field lacks a cohesive mathematical framework unifying EPs and TPDs across symmetry classes, systematic figures of merit for comparing TPD configurations, and design principles for selecting optimal operating points. 

\begin{figure}
\centering
\includegraphics[width=.44\textwidth]{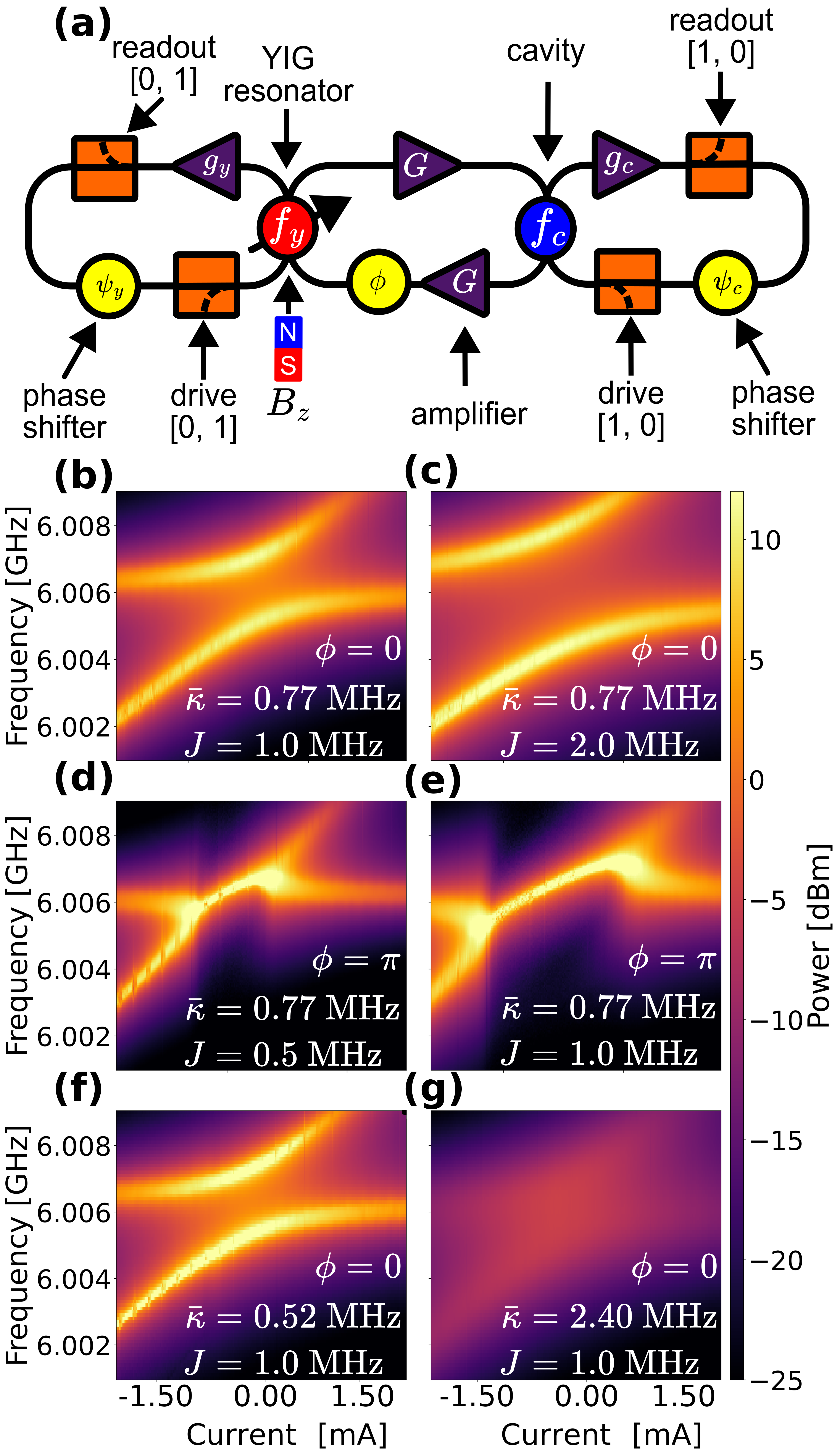}
\caption{(a) Schematic of our architecture. The coupling amplifier and phase shifter ($G, \phi$) determine the coupling rate ($J$) and phase ($\phi$). Auxiliary controls ($g_c, g_y, \psi_c, \psi_y, B_z$) modify the mode parameters $f_c, f_y, \kappa_c, \kappa_y$. (b)-(g) Transmission spectra as a function of current through an external electromagnet, which tunes the YIG frequency $f_y$ through $B_z$ and thereby controls the detuning between the modes. Data illustrate level repulsion ($\phi = 0$) (b),(c),(f),(g) and level attraction ($\phi = \pi$) (d),(e) for varying coupling strengths $J$ and average dissipation rates $\bar{\kappa}$.}
\label{fig:platform}
\end{figure}

Here, we establish a unified semiclassical framework for two-dimensional EPs and TPDs. In Sec. \ref{sec:arch} we implement a loop-coupled cavity-magnonic architecture (Fig.~\ref{fig:platform}) with the extensive parameter control necessary to navigate the non-Hermitian landscape. In Sec. \ref{sec:dynamics}, we map the parameter landscape of EPs and TPDs (Fig.~\ref{fig:theory}), establishing a general model for the dynamics, eigenspectrum, and transmission spectrum. We experimentally validate these predictions in Sec.~\ref{section:results}, systematically realizing six distinct EP–TPD configurations by varying the control parameters (Fig.~\ref{fig:eps}). With the parameter space mapped, Sec. \ref{section:design_guidance} derives analytic figures of merit to guide sensor design subject to noise, nuisance fluctuations, and experimental constraints (Fig.~\ref{fig:petermann}). Finally, in Sec. \ref{sec:ted}, we demonstrate that specific control parameters realize a robust TPD with minimal response to nuisance fluctuations (Fig.~\ref{fig:parametric}). Furthermore, we show that the third-order degeneracy of all TPDs offers a fundamental advantage for retaining sublinear splitting even in the presence of fluctuations. The formalism, platform, and experimental methods presented here establish a foundational reference for TPD-based sensor design while simultaneously presenting a versatile testbed for exploring non-Hermitian system dynamics. 

\section{\label{sec:arch} Tunable Architecture}

Our architecture is a magnon-photon dimer, consisting of the lowest frequency mode of a three-dimensional microwave cavity \cite{reagor_quantum_2016} with frequency $f_c$ and loaded full-width-half-maximum $\kappa_c$ and a magnon mode realized using a yttrium iron garnet (YIG) sphere of radius 1~mm. The magnon Larmor frequency is $f_y = \mu_B B_z/2\pi$, where $\mu_B$ is the Bohr magneton, and $B_z$ is the strength of an applied static magnetic field \cite{sparks_ferromagnetic_1961}. Two orthogonal wire bonds inductively couple to the magnon mode, passing signals around $f_y$, with loaded full-width-half-maximum $\kappa_y$, exhibiting a Lorentzian transmission spectrum \cite{barry_ferrimagnetic_2023}. 

\begin{figure}
\centering
\includegraphics[width=0.48\textwidth]
{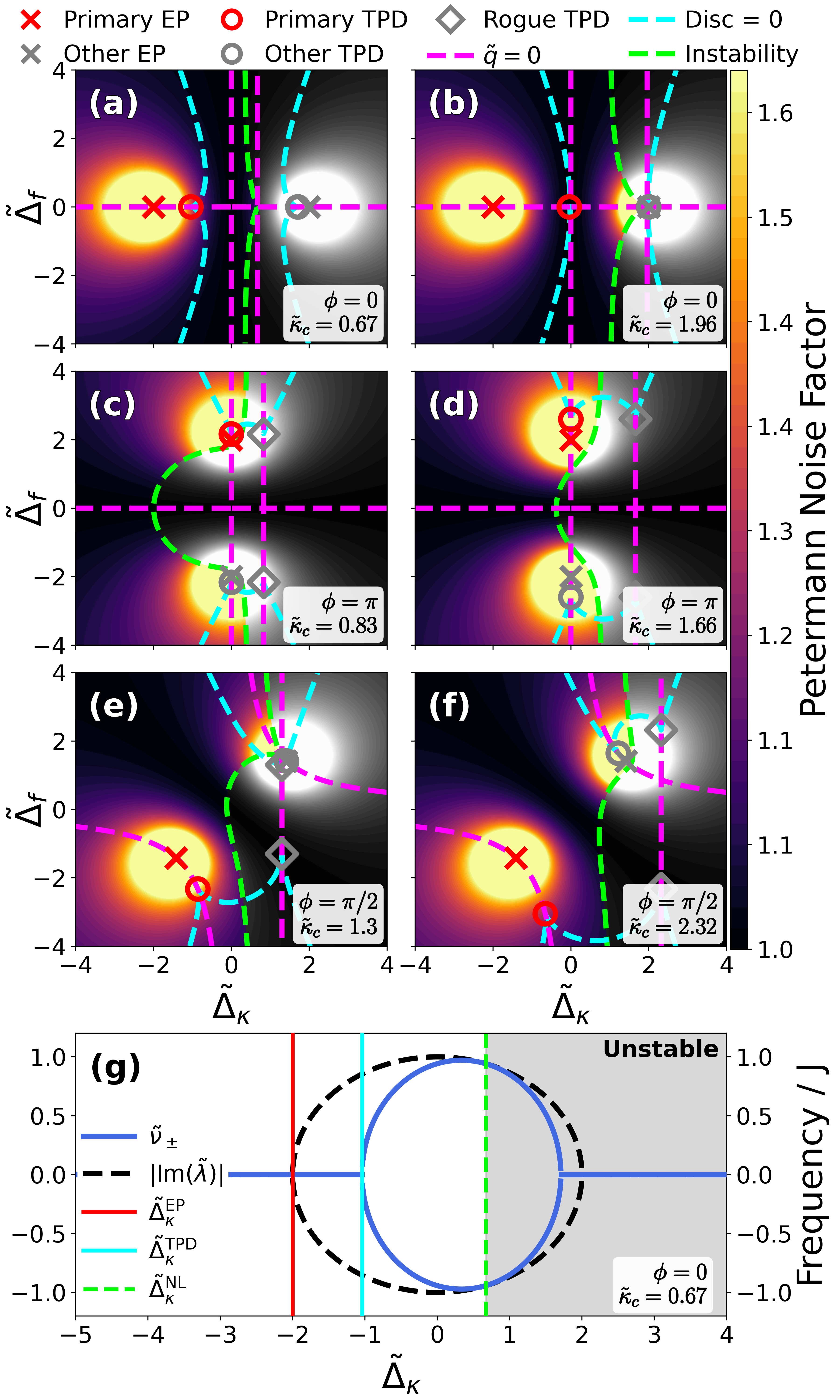}
\caption{Parameter landscape for $\phi = 0$ (a),(b), $\phi = \pi$ (c),(d), and $\phi = \pi/2$ (e),(f) overlaid atop the Petermann factor (clipped to 90th percentile). Each panel corresponds to the same $\phi$ and $\tilde{\kappa}_c$ used in the experimental panel in Fig.~\ref{fig:eps}. Red crosses (circles) are EPs (TPDs) experimentally implemented in Fig.~\ref{fig:eps}; ``other" EPs and TPDs are not considered here. Gray diamonds represent TPDs not associated with an EP, referred to as rogue TPDs. Cyan contours are $\mathrm{Disc} = 0$ (Eq.~\ref{eq:disc_algebraic}), separating the parameter space into single peak (outside cyan contours horizontally) and two peak (between cyan contours horizontally) regions. Magenta contours are $\tilde{q} = 0$ (Eq.~\ref{eq:q}), used to determine the independent variable in Fig.~\ref{fig:eps}. Lime contours are $\operatorname{Re}(\tilde{\Delta}_\lambda) = (\tilde{\kappa}_c - \tilde{\Delta}_\kappa)$, and separate the parameter space into stable (Inferno colorscale) and unstable (grayscale, linear model inapplicable) regimes. (g), an alternative representation of EPs and TPDs, representing a cross-section along the $\tilde{\Delta}_f$ axis for the $\tilde{\kappa}_c, \phi$ configuration in (a). The imaginary eigenvalues ($\operatorname{Im}(\tilde{\lambda})$) split at the EP (solid red vertical), while the transmission peak frequencies ($\tilde{\nu}$) split at the TPD (solid cyan vertical).}
\label{fig:theory}
\end{figure}

The cavity and YIG are spatially separated by approximately one meter and connected using Sub-Miniature version A (SMA) cables \cite{rao_meterscale_2023} (Fig.~\ref{fig:platform}(a)). We equip each coupling path with a tunable directional amplifier and one path with a digital phase shifter. Together, these elements symmetrically control the coupling strength $J>0$ and impart a controllable relative phase $\phi$, forming an effective synthetic gauge field \cite{salcedo-gallo_demonstration_2025, gardin_engineering_2024}. The resulting complex coupling coefficient, $Je^{i\phi}$, is tunable across the full range $\phi \in [0, 2\pi)$ while preserving the coupling magnitude, allowing this platform to be used to study $\mathcal{PT}$-symmetric ($\phi = 0$) \cite{bender_real_1998, regensburger_paritytime_2012}, anti-$\mathcal{PT}$-symmetric ($\phi = \pi$) \cite{peng_anti-paritytime_2016, li_antiparity-time_2019, choi_observation_2018, zhang_synthetic_2020}, and anyonic-$\mathcal{PT}$-symmetric ($\phi \notin \{0, \pi\}$) dynamics \cite{li_experimental_2024, arwas_anyonic-parity-time_2022, gao_parity-time-anyonic_2019}. At $\phi = 0$, the system exhibits level repulsion, characterized by an avoided crossing as $f_y$ is tuned through $f_c$, whose gap increases with $J$ \cite{harder_level_2018} (Fig.~\ref{fig:platform}(b),(c),(f),(g)). Modifying the average dissipation rate, $\bar{\kappa} = (\kappa_c + \kappa_y)/2$, changes the linewidths of the hybridized modes (Fig.~\ref{fig:platform}(f),(g)). At $\phi = \pi$, level attraction is observed, where the hybridized modes coalesce and split as the individual frequencies are detuned (Fig.~\ref{fig:platform}(d),(e)). As expected, the extent of the coalescence region increases with $J$. For all other phase values, the two couplings still share the same magnitude, but their complex phases differ by an angle that cannot be removed by a single phase redefinition or gauge transformation. The dynamics of these phase-non-reciprocal configurations are explored in Figs.~\ref{fig:theory}, \ref{fig:eps}, and \ref{fig:petermann}.

To complete the six-parameter control of the dimer, each mode is equipped with a self-feedback loop consisting of a digital phase shifter with phase $\psi_{c,y}$ and an effective variable-gain amplifier with gain $g_{c,y}$ (Fig.~\ref{fig:platform}(a)), where the subscripts $(c, y)$ correspond to the cavity and YIG self-feedback loops, respectively. For the YIG self-feedback loop, a voltage-controlled analog attenuator tunes $g_y$ instead, such that $\kappa_y$ is a function of voltage. Adjusting $\psi_{c,y}$ and $g_{c,y}$ provides in situ tuning of the mode frequencies $f_{c,y}$ and dissipation rates $\kappa_{c,y}$. Figure~\ref{fig:platform} illustrates the experimental schematic (Fig.~\ref{fig:platform}(a)) and various transmission experiments highlighting parameter control (Fig.~\ref{fig:platform}(b--g)).

\section{\label{sec:dynamics} Non-Hermitian Dynamics}

The dynamics of the system are described using a state-space model, governed by the semiclassical equations of motion in the rotating frame with respect to the normalized drive frequency $\tilde f_d$,
\begin{align}
\frac{d}{dt}\,\tilde{\boldsymbol{\alpha}} &= \tilde{\boldsymbol{A}}\,\tilde{\boldsymbol{\alpha}} + \boldsymbol{B}\,u, \label{eq:time_dep}\\
\tilde{\boldsymbol{\alpha}}_{\mathrm{ss}}(\tilde{f}_d) &= (i\,\tilde{f}_d\,\boldsymbol{I} + \tilde{\boldsymbol{A}})^{-1}\boldsymbol{B}\,u, \label{eq:alpha_ss}\\
\tilde{\beta}_{\mathrm{ss}}(\tilde{f}_d) &= \bigl|\boldsymbol{C}\,\tilde{\boldsymbol{\alpha}}_{\mathrm{ss}}(\tilde{f}_d)\bigr|^2, \label{eq:beta_ss}
\end{align}
where $\tilde{\boldsymbol{A}}$ is the nondimensionalized dynamical system matrix defined in Eq.~\ref{eq:model}, $\boldsymbol{I}$ is the identity matrix, and $\tilde{\boldsymbol{\alpha}} = [\tilde{\alpha}_c, \tilde{\alpha}_y]^{\mathrm T}$ denotes the complex normalized coherent state amplitudes of the photon ($\tilde{\alpha}_c$) and magnon ($\tilde{\alpha}_y$) modes, and $\tilde{\boldsymbol{\alpha}}_{\mathrm{ss}}$ denotes the corresponding steady-state solution. $\boldsymbol{B} \in \{[1,0]^{\mathrm T}, [0,1]^{\mathrm T}\}$ is the drive vector, where $\boldsymbol{B} = [1,0]^{\mathrm T}$ ($\boldsymbol{B} = [0,1]^{\mathrm T}$) represents driving the cavity (YIG) with drive strength $u$. Complex mode amplitudes are measured by $\boldsymbol{C}\tilde{\boldsymbol{\alpha}}$ with $\boldsymbol{C} \in \{[1,0], [0,1]\}$ the readout vector, $\boldsymbol{C} = [1,0]$ ($\boldsymbol{C} = [0,1]$) corresponds to reading out the cavity (YIG). $\tilde{\beta}_{\mathrm{ss}}$ in Eq.~\ref{eq:beta_ss} represents the steady state transmission spectrum, which is only valid when all real parts of the eigenvalues of $\tilde{\boldsymbol{A}}$ are negative. When $\tilde{\boldsymbol{A}}$ becomes unstable, amplifier saturation leads to nonlinear phenomena such as synchronization, nonlinear exceptional points, and generalized $\mathcal{PT}$ symmetry \cite{bai_nonlinear_2023, salcedo-gallo_demonstration_2025, wu_generalized_2022}.

The system matrix generates dynamics and is given in nondimensionalized form by
\begin{align}\label{eq:model}
\tilde{\boldsymbol{A}} =  \begin{bmatrix}
-i\tilde{f}_c - \frac{\tilde{\kappa}_c}{2}  & -i \\ 
-i e^{i\phi} & -i(\tilde{f}_c  - \tilde{\Delta}_f) + \left(\tilde{\Delta}_\kappa - \frac{\tilde{\kappa}_c}{2}\right)
\end{bmatrix}
\end{align}
where $\tilde{f}_c = f_c/J$, $\tilde{\kappa}_c = \kappa_c/J$, etc., and $\tilde{\Delta}_f = \tilde{f}_c - \tilde{f}_y$, and $\tilde{\Delta}_\kappa = (\tilde{\kappa}_c - \tilde{\kappa}_y)/2$ (see Appendix \ref{appendix:a} for derivation) \cite{yao_microscopic_2019}. 

\subsection{\label{subsec:eigenvalues} Eigenspectrum Analysis}

The eigenvalues \( \tilde{\lambda}_\pm \) of \( \tilde{\boldsymbol{A}} \) are given by 
\begin{align}
\tilde{\Delta}_\lambda &\equiv \sqrt{ (\tilde{\Delta}_\kappa + i\tilde{\Delta}_f)^2 - 4 e^{i\phi} }, \label{eq:deltalambda_tilde} \\
\tilde{\lambda}_\pm &= \left( \frac{\tilde{\Delta}_\kappa}{2} - \frac{\tilde{\kappa}_c}{2} \right) + i \left( \frac{\tilde{\Delta}_f}{2} - \tilde{f}_c \right) \pm \frac{\tilde{\Delta}_\lambda}{2}, \label{eq:lambda0_tilde}
\end{align}
where \( \operatorname{Re}(\tilde{\lambda}) \) is related to the dissipation or gain, and \( \operatorname{Im}(\tilde{\lambda}) \) is related to the oscillation frequency of each mode. This convention is opposite to the typical Hamiltonian-based formulation, but provides consistent results while offering a direct connection to electrical engineering \cite{salcedo-gallo_demonstration_2025} and modern control theory \cite{juang_identification_2001}. 

EPs are defined by the simultaneous coalescence of eigenvectors and eigenvalues. We locate EPs using the mean diagonal Petermann factor, which equals unity for orthogonal eigenvectors and diverges at EPs (see Appendix \ref{appendix-sec:petermann} for additional information) \cite{zheng_mathcalpt_2010, ashida_non-hermitian_2020, siegman_excess_1989}. For $\tilde{\boldsymbol{A}}$ in Eq.~\ref{eq:model}, we find
\begin{align}
    \text{PF} = \frac{\tilde{\Delta}_f^2 + \tilde{\Delta}_\kappa^2 + |\tilde{\Delta}_\lambda|^2 + 4}{2|\tilde{\Delta}_\lambda|^2},
\end{align}
which diverges as $\tilde{\Delta}_\lambda \rightarrow 0$. This divergence condition reveals a continuous manifold of EPs in parameter space, a so-called exceptional surface \cite{zhou_exceptional_2019}. We define EP locations as $\boldsymbol{\tilde{\Delta}}^\mathrm{EP} \equiv (\tilde{\Delta}_\kappa^\text{EP}, \tilde{\Delta}_f^\text{EP})$ given by
\begin{align}
    \tilde{\Delta}_\kappa^\text{EP} = \pm 2\cos\bigg(\frac{\phi}{2}\bigg), \quad 
    \tilde{\Delta}_f^\text{EP} = \pm 2\sin \bigg(\frac{\phi}{2}\bigg).
\end{align}
At each value of $\phi$, there are two EPs (``primary" and ``other"), depicted in Fig.~\ref{fig:theory}, with ``primary" EPs (red crosses) investigated experimentally in Fig.~\ref{fig:eps}. Near EPs, the diverging Petermann factor corresponds to noise enhancement \cite{wang_petermann-factor_2020, ghosh_suppression_2024}. Moving away from the EPs in $\boldsymbol{\tilde{\Delta}}$-space reduces the Petermann factor, reaching unity between the two EPs on a line parameterized by $\tilde{\Delta}_f = -\cot(\phi/2) \tilde{\Delta}_\kappa$.  

\subsection{\label{subsec:transmission_spectrum} Transmission Spectrum Analysis}

We obtain transmission frequency extrema by solving $\frac{\partial}{\partial \tilde{f}_d}(\tilde{\beta}_{\mathrm{ss}})=0$ for $\tilde{f}_d$, where the denominator of $\tilde{\beta}_{\mathrm{ss}}$ is a quartic polynomial in $\tilde f_d$. Thus, the locations of the extrema of $\tilde{\beta}_{\mathrm{ss}}$ with respect to $\tilde{f}_d$ depend on a cubic-order polynomial. 
We simplify the extrema condition into a depressed cubic equation with auxiliary variable $x$ of the form
\begin{align}
    0 &= x^3 + \tilde{p}x + \tilde{q}, \label{eq:cubic} \\
    \tilde{p} &= \frac{ (\tilde{\kappa}_c - \tilde{\Delta}_\kappa)^2 + \operatorname{Re}(\tilde{\Delta}_\lambda^2) }{4}, \label{eq:p} \\
    \tilde{q} &= \frac{ (\tilde{\kappa}_c - \tilde{\Delta}_\kappa) \, \operatorname{Im}(\tilde{\Delta}_\lambda^2) }{8}. \label{eq:q}
\end{align}
This formulation distills the extrema condition into fundamental physical parameters: the eigenvalue splitting $\tilde{\Delta}_\lambda$ (Eq. \ref{eq:deltalambda_tilde}), and $\tilde{\kappa}_c - \tilde{\Delta}_\kappa$, which can be identified as the normalized average dissipation rate $(\tilde{\kappa}_c + \tilde{\kappa}_y)/2$. 

The discriminant ($\mathrm{Disc}$) governs the nature of the roots of Eq.~\ref{eq:cubic}, where
\begin{align} 
\mathrm{Disc} &= -4\tilde p^3 - 27 \tilde q ^2, \label{eq:disc_algebraic}\\ &= (\tilde \nu_+^\text{Root} \!\!-\! \tilde \nu_-^\text{Root})^2 (\tilde \nu_+^\text{Root} \!\!-\!  \tilde \eta^\text{Root})^2 (\tilde \nu_-^\text{Root} \!\!-\!  \tilde \eta^\text{Root})^2. \nonumber
\end{align}
Equation~\ref{eq:disc_algebraic} gives the algebraic form of the discriminant, or equivalently, in terms of the roots. When $\mathrm{Disc} < 0$, Eq.~\ref{eq:cubic} admits one real root, $\tilde{\nu}_0^{\text{Root}}$, and the system exhibits a single transmission peak located at $\tilde{\nu}_0 = \tilde{\nu}_0^{\text{Root}}+\tilde{f}_c-\tilde{\Delta}_f/2$. When $\mathrm{Disc} > 0$, Eq.~\ref{eq:cubic} admits three real roots, $\tilde{\nu}_\pm^{\text{Root}}, \tilde{\eta}^{\text{Root}}$. In this case, $\tilde{\nu}_\pm = \tilde{\nu}_\pm^\text{Root} + \tilde{f}_c - \frac{\tilde{\Delta}_f}{2}$ are the frequencies of the two transmission peaks, and $\tilde{\eta} = \tilde{\eta}^\text{Root} + \tilde{f}_c - \frac{\tilde{\Delta}_f}{2}$ is the frequency of the local minimum between the two peaks. Solving Eq.~\ref{eq:cubic} for a closed form solution of the roots in terms of $\tilde{p}, \tilde{q}$ can either be achieved numerically, using Cardano's method, or through a trigonometric solution first posed by Fran\c{c}ois Vi\`ete in 1615 \cite{nickalls_viete_2006, viete_ad_1615}, with
\begin{align}
    \tilde{\theta} &= \frac{1}{3}\arccos\!\left(
        \frac{3\tilde{q}}{2\tilde{p}}\sqrt{\frac{-3}{\tilde{p}}}
    \right), \label{eq:theta} \\
    \tilde{\nu}_+^\text{Root} &= 2\sqrt{\frac{-\tilde{p}}{3}}\,
        \cos\!\left(\tilde{\theta}\right), \label{eq:nu_plus} \\
    \tilde{\eta}^{\text{Root}} &= 2\sqrt{\frac{-\tilde{p}}{3}}\,
        \cos\!\left(\tilde{\theta} - \frac{2\pi}{3}\right), \label{eq:eta} \\
    \tilde{\nu}_-^{\text{Root}} &= 2\sqrt{\frac{-\tilde{p}}{3}}\,
        \cos\!\left(\tilde{\theta} - \frac{4\pi}{3}\right). \label{eq:nu_minus}
\end{align}
The transmission peak splitting is $\tilde{\Delta}_\nu \equiv |\tilde{\nu}_+ - \tilde{\nu}_-|$, analogous to $\tilde{\Delta}_\lambda$ from Eq.~\ref{eq:deltalambda_tilde}. Equations~\ref{eq:theta}-\ref{eq:nu_minus} are valid when $\mathrm{Disc} > 0$. We give the closed-form solution for $\tilde{\nu}_0^\text{Root}$ in the single-peak regime ($\mathrm{Disc} < 0$) in Appendix \ref{appendix-sec:single_root}. 

\begin{figure}
\centering
\includegraphics[width=0.44\textwidth]{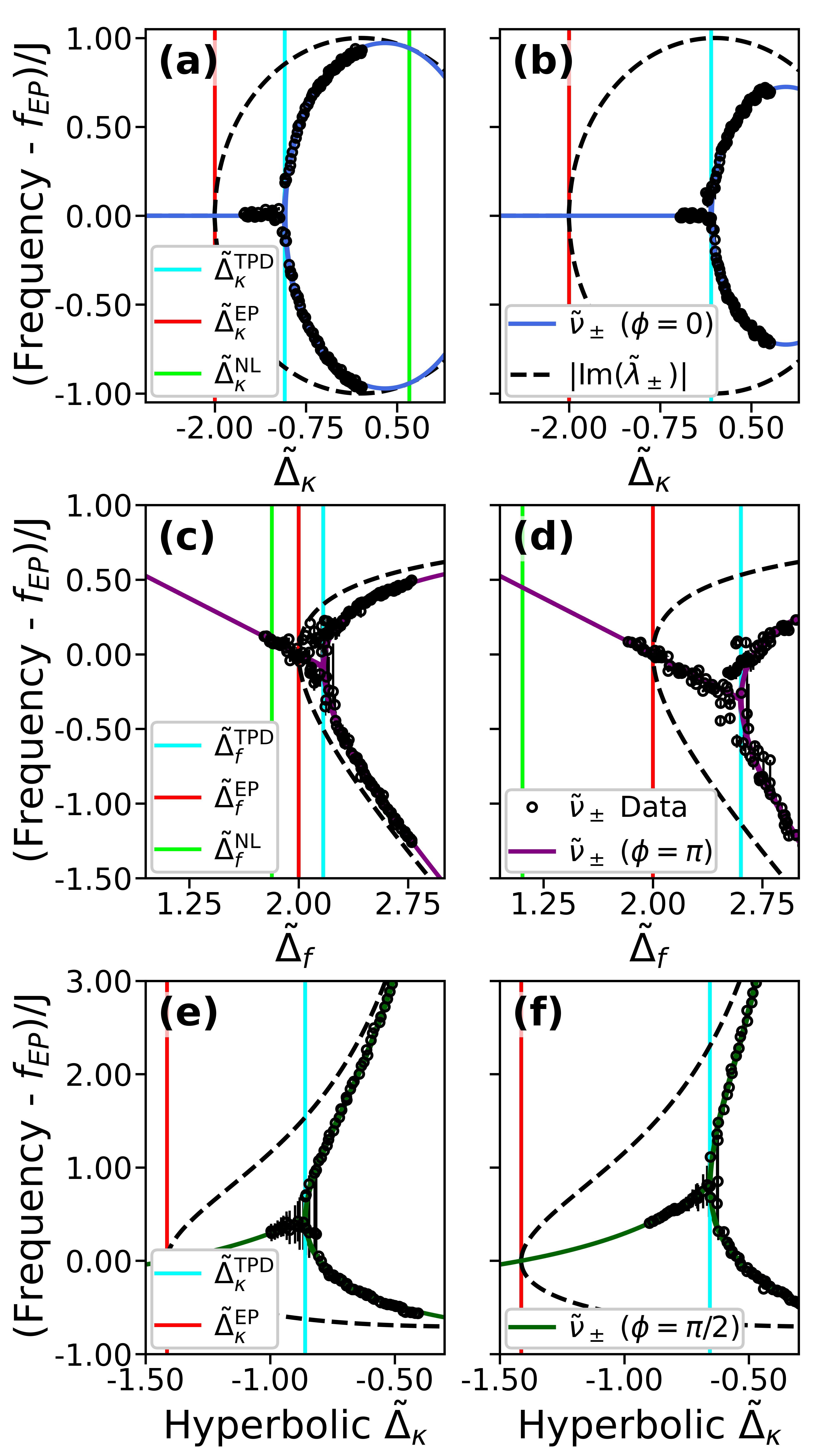}
\caption{Experimentally fit transmission peak locations $\tilde{\nu}_\pm$ overlaid atop theory (solid lines) and $|\operatorname{Im}(\tilde{\lambda}_\pm)|$ (dashed lines). The coupling phase ($\phi$) determines the relevant sweep parameter. (a),(b) $\phi = 0$ (blue) sweeps $\tilde \Delta_\kappa$, (c),(d) $\phi =\pi$ (purple) sweeps $\tilde \Delta_f$, and (e),(f) $\phi = \pi/2$ (green) sweeps along a hyperbolic path ($\tilde \Delta_\kappa \tilde \Delta_f = 2$) projected onto the $\tilde \Delta_\kappa$ axis. In all cases, the TPD location moves with $\tilde{\kappa}_c$, while the EP is static, with the left (right) column surveying smaller (larger) $\tilde{\kappa}_c$ values that place the TPD and EP closer to (further from) each other.}
\label{fig:eps}
\end{figure}

Naturally, the condition $\mathrm{Disc} = 0$ marks the boundary between the single-peak and split-peak regimes. From the product definition of the discriminant in Eq.~\ref{eq:disc_algebraic}, we observe that $\mathrm{Disc} = 0$ indicates a root degeneracy, but not necessarily the triple degeneracy required for a TPD. A standard root coalescence where a peak $\tilde{\nu}_\pm^\text{Root}$ merges with the minimum  $\tilde{\eta}^\text{Root}$ is referred to as a transmission extrema degeneracy (TED). TPDs are distinguished from TEDs by solving $\mathrm{Disc}=0$ and $\tilde{q}=0$ simultaneously. Figure~\ref{fig:theory} depicts the contours $\mathrm{Disc}=0$ (cyan) and $\tilde{q}=0$ (magenta); their intersections mark the TPDs. For each $\phi$ there are two (Fig.~\ref{fig:theory}(a),(b)), four (Fig.~\ref{fig:theory}(c)--(f)), or six TPDs, with locations $\boldsymbol{\tilde{\Delta}}^\text{TPD} \equiv (\tilde{\Delta}_\kappa^\text{TPD}, \tilde{\Delta}_f^\text{TPD})$ dependent on $\tilde{\kappa}_c$. We classify them as primary TPDs (red circles, experimentally validated in Fig.~\ref{fig:eps}, and the focus of this work), secondary TPDs (gray circles), and rogue TPDs (gray diamonds, detached from any EP). For fixed $\phi$, the EPs are stationary while TPDs can be moved with $\tilde{\kappa}_c$. The full parameterization of the TPD surface is given in Appendix \ref{appendix:b}.

\section{\label{section:results} Experimental Validation}

Here, we experimentally validate the square-root splitting of transmission peaks at TPDs by probing along the $\tilde{q} = 0$ contour that intersects both the primary EP and TPD. This trajectory enforces $\operatorname{Re}(\tilde{\Delta}_\lambda)=0$ along the sweep, ensuring that the peaks split with equal amplitude. For $\phi = 0$, the $\tilde{q} = 0$ path is the $\tilde{\Delta}_\kappa$ axis (Fig.~\ref{fig:theory}(a),(b)), tuned experimentally by varying the voltage applied to an analog attenuator. For $\phi = \pi$, the path is the $\tilde{\Delta}_f$ axis (Fig.~\ref{fig:theory}(c),(d)), controlled by the magnetic field applied to the YIG ($B_z$). For intermediate values of $\phi$, the path follows a hyperbola defined by $\tilde{\Delta}_\kappa \tilde{\Delta}_f = 2 \sin(\phi)$ (Fig.~\ref{fig:theory}(e),(f)), requiring simultaneous control of $\tilde{\Delta}_f$ and $\tilde{\Delta}_\kappa$. Arbitrary paths in $\boldsymbol{\tilde{\Delta}}$-space are tracked by our experimental apparatus using a parameter feedback control system (Appendix \ref{appendix-sec:methods-control}).

Our experimental methods are summarized here and detailed in Methods (Appendix \ref{appendix:d}). For each $\phi$, we calibrate $J$ to convert measured $\Delta_f, \Delta_\kappa$ to their dimensionless forms. Sweeps are initialized by calibrating $\tilde{\kappa}_c$ and defining a $\tilde{\Delta}_f$ or $\tilde{\Delta}_\kappa$ range to sweep. The experiment tracks $\tilde{q} = 0$ in $\boldsymbol{\tilde{\Delta}}$-space. At each step, we first probe the uncoupled cavity (YIG) modes ($J=0$), fit them to Lorentzians to extract $f_{c(y)}, \kappa_{c(y)}$, and propagate uncertainties $\sigma(f_{c(y)}), \sigma(\kappa_{c(y)})$ to $\tilde{\Delta}_f, \tilde{\Delta}_\kappa$. We then enable coupling ($J>0$), record transmission through the cavity to the YIG mode, and fit Eq.~\ref{eq:beta_ss}, extracting peak locations $\tilde{\nu}_\pm$ from the roots of Eq.~\ref{eq:cubic}. Error bars represent uncertainty propagation from $\sigma(\tilde{\Delta}_f), \sigma(\tilde{\Delta}_\kappa)$.

The locations of EPs and TPDs along the $\tilde{q} = 0$ trajectory are visualized in Fig.~\ref{fig:eps}. For $\phi = 0$, we calibrate $J=1.05(1)$~MHz, and scan $\tilde{\Delta}_\kappa$ while keeping $\tilde{\Delta}_f = 0$. This reveals a static EP at $\tilde{\Delta}_\kappa^\text{EP}=\pm 2$ and a movable TPD at $\tilde{\Delta}_\kappa^\text{TPD} = \frac{1}{2}(\tilde{\kappa}_c \pm \sqrt{8 - \tilde{\kappa}_c^2})$. Figure~\ref{fig:eps}(a),(b) confirms this behavior, depicting the TPD at $\tilde{\Delta}_\kappa^\text{TPD} = -1.04(1)$ for $\tilde{\kappa}_c = 0.67(1)$ (Fig.~\ref{fig:eps}(a)) and $\tilde{\Delta}_\kappa^\text{TPD} = -0.05(4)$ for $\tilde{\kappa}_c = 1.95(1)$ (Fig.~\ref{fig:eps}(b)). In both cases, the transmission peaks $\tilde{\nu}_\pm$ split at the TPD while remaining bounded by $|\text{Im}(\tilde{\lambda}_\pm)|$. The linear regime is bounded by the instability transition observed at $\tilde{\Delta}_\kappa^\text{NL} = 0.67(1)$ (Fig.~\ref{fig:eps}(a)), consistent with the theoretical predictions $\tilde{\Delta}_\kappa^\text{NL} = \min(\tilde{\kappa}_c, (\tilde{\kappa}_c^2 + 4)/2\tilde{\kappa}_c)$.

For $\phi = \pi$, we calibrate $J=1.12(2)$~MHz, and scan $\tilde{\Delta}_f$ while keeping $\tilde{\Delta}_\kappa = 0$. This reveals static EPs at $\tilde{\Delta}_f^\text{EP} = \pm 2$, movable TPDs at $\tilde{\Delta}_f^\text{TPD} = \pm \sqrt{\tilde{\kappa}_c^2 + 4}$, and instability transitions at $\tilde{\Delta}_f^\text{NL} = \pm \sqrt{4 - \tilde{\kappa}_c^2}$. Figure~\ref{fig:eps}(c),(d) depicts the TPD and instability transition at $\tilde{\Delta}_f^\text{TPD} = 2.168(4)$, $\tilde{\Delta}_f^\text{NL} = 1.817(4)$ for $\tilde{\kappa}_c = 0.84(1)$ (Fig.~\ref{fig:eps}(c)) and $\tilde{\Delta}_f^\text{TPD} = 2.603(7)$, $\tilde{\Delta}_f^\text{NL} = 1.117(7)$ for $\tilde{\kappa}_c = 1.67(1)$ (Fig.~\ref{fig:eps}(d)). 

For $\phi = \pi/2$, we calibrate $J=0.95(2)$~MHz and scan along the hyperbolic trajectory $\tilde{\Delta}_f = 2/\tilde{\Delta}_\kappa$ to maintain $\tilde{q} = 0$. This reveals a static EP at $\boldsymbol{\tilde{\Delta}}^\text{EP} = (-\sqrt{2}, -\sqrt{2})$ and movable TPDs without a reasonable closed form solution (Appendix \ref{appendix:b}). Figure~\ref{fig:eps}(e),(f) depicts the TPD (projected onto the $\tilde{\Delta}_\kappa$ axis) at $\boldsymbol{\tilde{\Delta}}^\text{TPD} = (-0.860(3), -2.327(9))$ for $\tilde{\kappa}_c = 1.30(1)$ (Fig.~\ref{fig:eps}(e)) and $\boldsymbol{\tilde{\Delta}}^\text{TPD} = (-0.657(4), -3.04(2))$ for $\tilde{\kappa}_c = 2.32(3)$ (Fig.~\ref{fig:eps}(f)). Unlike the $\phi = 0, \pi$ configurations, the instability transition remains inaccessible within the experimental window.

\begin{figure*}[htbp!]
\begin{centering}
\includegraphics[width=0.98\textwidth]{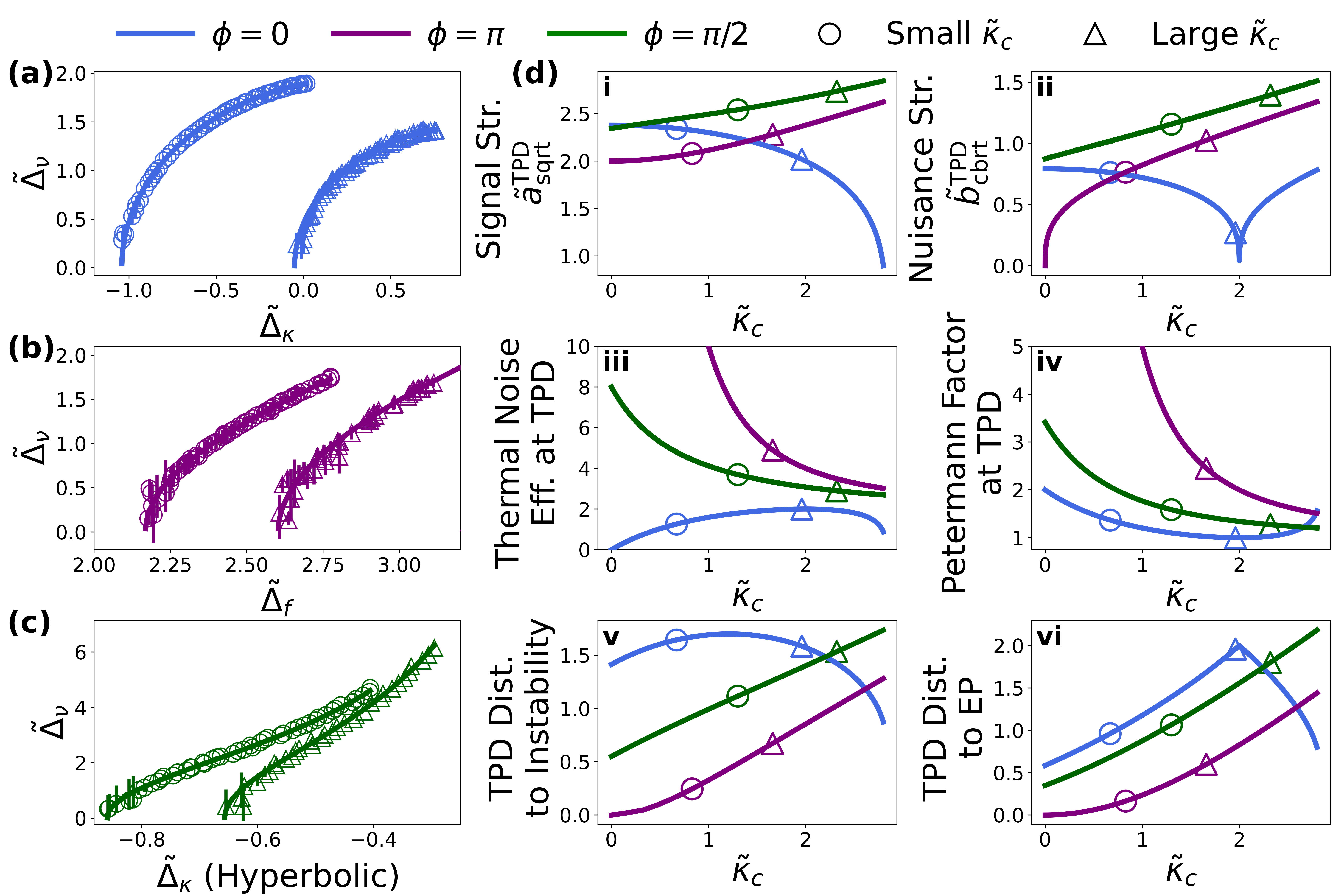}
\end{centering}
\caption{(a)–(c) Experimental data from Fig.~\ref{fig:eps} recast as peak splitting $\tilde{\Delta}_\nu$, versus the sweep parameter for $\phi = 0$ (a), $\phi = \pi$ (b), and $\phi = \pi/2$ (c). Markers distinguish small $\tilde{\kappa}_c$ (circles, data from left column from Fig.~\ref{fig:eps}) and large $\tilde{\kappa}_c$ (triangles, data from right column from Fig.~\ref{fig:eps}) configurations. (d) Semiclassical sensing figures of merit as functions of $\tilde{\kappa}_c$ for $\phi = 0$ (blue), $\phi = \pi$ (purple) and $\phi = \pi/2$ (green). (d.i) target signal splitting strength $\tilde{a}_{\text{sqrt}}^\text{TPD}$, (d.ii) nuisance scaling $\tilde{b}_{\text{cbrt}}^\text{TPD}$, (d.iii) thermal noise efficiency at the TPD, (d.iv) Petermann factor at the TPD, (d.v) distance from the TPD to the nearest instability transition, and (d.vi) distance to the nearest EP. Markers indicate the experimental configurations from (a)–(c).}
\label{fig:petermann}
\end{figure*}

\section{\label{section:design_guidance} Figures of Merit for Sensor Design}
In Sec. \ref{section:results}, we establish that TPDs can be moved in $\boldsymbol{\tilde{\Delta}}$-space using the control parameters $\phi$ and $\tilde{\kappa}_c$. Here, we examine the practical implications of this tunability for sensing applications, providing the guidance necessary to select the optimal TPD candidate based on specific experimental and noise constraints.

Fundamentally, the sensing protocol used in Refs. \cite{kononchuk_exceptional-point-based_2022, lu_harnessing_2025} exploits the symmetric square-root splitting of transmission peaks (Fig.~\ref{fig:eps}) to detect external perturbations. We define the target signal perturbation $\boldsymbol{\tilde{\epsilon}} \equiv (\tilde{\epsilon}(\tilde{\Delta}_\kappa), \tilde{\epsilon}(\tilde{\Delta}_f)) > 0$ as a parameter shift at the TPD strictly aligned with the optimal sensing trajectory ($\tilde q = 0$) established in Sec. \ref{section:results}. Along this trajectory, the peak splitting scales as
\begin{align}
\tilde{\Delta}_\nu(\boldsymbol{\tilde{\Delta}}^\text{TPD}+\boldsymbol{\tilde{\epsilon}}) \approx  \tilde{a}_\text{sqrt}^\text{TPD} \sqrt{\lVert \boldsymbol{\tilde{\epsilon}} \rVert}, \label{eq:splitting_expansion_target}
\end{align}
where $\tilde{a}_\text{sqrt}^\text{TPD}$ is a scaling coefficient defined in Sec. \ref{subsec:signal_scaling}.  

By strictly defining the signal perturbation $\boldsymbol{\tilde{\epsilon}}$ as the component along this optimal trajectory, we distinguish it from nuisance perturbations $\boldsymbol{\tilde{\delta}} \equiv (\tilde{\delta}(\tilde{\Delta}_\kappa), \tilde{\delta}(\tilde{\Delta}_f))$, which are orthogonal drifts that force the system off this path with $\boldsymbol{\tilde{\delta}} \cdot \boldsymbol{\tilde{\epsilon}} = 0$. The sensor response to the target signal is described using the susceptibility $\boldsymbol{\tilde{\chi}}_{\boldsymbol{\tilde{\epsilon}}} = \nabla_{\boldsymbol{\tilde{\epsilon}}} \tilde{\Delta}_\nu$, which diverges as $\tilde{a}_\text{sqrt}^\text{TPD}/(2\sqrt{\lVert\boldsymbol{\tilde{\epsilon}}\rVert})$ at TPDs (where $\lVert\boldsymbol{\tilde{\epsilon}}\rVert \to 0$), reminiscent of the behavior at EPs \cite{wiersig_enhancing_2014}.

While the divergence of $\boldsymbol{\tilde{\chi}}_{\boldsymbol{\tilde{\epsilon}}}$ at the TPD is a universal feature, the practical utility of a TPD sensor is influenced by specific figures of merit. Here, we consider figures of merit that quantify the scaling coefficient of the peak splitting, the amplification of thermal noise, and the robustness against nuisance drift. These figures of merit are dependent on $\phi$ and $\tilde \kappa_c$, allowing the optimal TPD configuration to be selected based on the constraints of specific sensing implementations 

\subsection{\label{subsec:signal_scaling}Scaling of Frequency Peak Splitting}

Here, we provide the explicit forms of the scaling coefficient $\tilde{a}_\text{sqrt}^\text{TPD}$ introduced in Eq.~\ref{eq:splitting_expansion_target}. A similar metric was recently reported for the $\phi=\pi$ TPD \cite{kononchuk_exceptional-point-based_2022}. Here, we formalize the definition for all TPDs by analyzing the (first order) Puiseux series expansion of $\tilde{\Delta}_\nu$ along the target signal trajectories from Sec. \ref{section:results}, and find
\begin{align}
\tilde{\Delta}_\nu(\phi = 0) &\approx
\sqrt{2}\left(8-\tilde{\kappa}_{c}^{2}\right)^{1/4}
\sqrt{\tilde{\epsilon}(\tilde{\Delta}_\kappa)},
\label{eq:phi_0_tpd_strength} \\
\tilde{\Delta}_\nu(\phi=\pi) &\approx \sqrt{2}\left( \tilde{\kappa}_c^2 + 4
\right)^{1/4} 
\sqrt{\tilde{\epsilon}(\tilde{\Delta}_f)} , 
\label{eq:phi_pi_tpd_strength}
\end{align}
where the pre-factor to the square-root term is $\tilde{a}_{\text{sqrt}}^\text{TPD}$. Equation~\ref{eq:phi_0_tpd_strength} is valid for $0 < \tilde{\epsilon}(\tilde{\Delta}_\kappa) < \sqrt{8 - \tilde{\kappa}_c^2}$. Equation~\ref{eq:phi_pi_tpd_strength} is valid for $\tilde{\epsilon}(\tilde{\Delta}_f)>0$. For $\phi = \pi/2$, $\tilde{a}_{\text{sqrt}}^\text{TPD}$ is not reported in closed form, but can be computed numerically. Figure~\ref{fig:petermann}(d.i) displays $\tilde{a}_{\text{sqrt}}^\text{TPD}$ as a function of $\tilde{\kappa}_c$ for $\phi = 0, \pi, \pi/2$. 

\subsection{\label{subsec:additive_noise}Semiclassical Petermann and Thermal Noise}

While $\boldsymbol{\tilde{\chi}}$ and $\tilde{a}_{\text{sqrt}}^\text{TPD}$ govern the signal response, the ultimate sensitivity performance is constrained by the SNR. In the semiclassical limit, additive noise arises primarily from thermal fluctuations. As discussed in Sec. \ref{subsec:eigenvalues}, the Petermann factor quantifies the noise amplification resulting from eigenbasis collapse. While the Petermann factor diverges at EPs, TPDs generally retain a complete eigenbasis and a finite Petermann factor (Fig.~\ref{fig:petermann}(d.iv)). Evaluating the Petermann factor at $\boldsymbol{\tilde{\Delta}} = \boldsymbol{\tilde{\Delta}}^\text{TPD}$ yields
\begin{align}
    \text{PF}^\text{TPD}(\phi = 0) &= \frac{8}{\tilde{\kappa}_c \sqrt{8 - \tilde{\kappa}_c^2}+4},\label{pf_tpd_phi_0} \\
    \text{PF}^\text{TPD}(\phi = \pi) &= \frac{{\tilde{\kappa}_c}^2 +4}{{\tilde{\kappa}_c}^2},\label{pf_tpd_phi_pi}
\end{align}
where Eq. \ref{pf_tpd_phi_0} is valid for $\tilde{\kappa}_c \leq 2\sqrt{2}$. Figure~\ref{fig:petermann}(d.iv) plots Eqs. \ref{pf_tpd_phi_0}-\ref{pf_tpd_phi_pi} alongside the numerically calculated Petermann factor for $\phi = \pi/2$. The Petermann factor at a TPD is related to the Euclidean distance in $\boldsymbol{\tilde{\Delta}}$-space between the TPD and nearest EP (Fig.~\ref{fig:petermann}(d.vi)), defined as $\text{dist}_\text{EP}(\phi, \tilde{\kappa}_c) \equiv \min_\text{EP} \lVert \boldsymbol{\tilde{\Delta}}^\text{TPD}(\phi, \tilde{\kappa}_c) - \boldsymbol{\tilde{\Delta}}^\text{EP}(\phi) \rVert _2$. As $\text{dist}_\text{EP} \to 0$, the TPD and EP coalesce, and the Petermann factor diverges.

For TPDs where the Petermann factor is finite, a more representative metric for additive noise performance is the thermal noise efficiency ($\mathrm{N}_\text{th}$). This dimensionless metric quantifies the transduction of uncorrelated thermal fluctuations from the system's reservoirs to the readout port, normalized by $4k_B T_\text{eff} f_{\text{bw}}$, where $T_\text{eff}$ is the effective noise temperature, and $f_{\text{bw}}$ is the measurement resolution bandwidth. In terms of the system Green's function $\boldsymbol{\tilde{G}} = (i\tilde{f}_d \boldsymbol{I} + \boldsymbol{\tilde{A}})^{-1}$, the thermal noise efficiency is \cite{kononchuk_exceptional-point-based_2022, clerk_introduction_2010}
\begin{align}
    \mathrm{N}_\text{th} &\equiv \tilde{\kappa}_y^2 |\boldsymbol{\tilde{G}}_{2,2}|^2 + \tilde{\kappa}_c\tilde{\kappa}_y|\boldsymbol{\tilde{G}}_{2,1}|^2.
\end{align}
We evaluate $\mathrm{N}_\text{th}$ at the TPD parameters ($\tilde{f}_d = \tilde{f}_d^\text{TPD}, \boldsymbol{\tilde{\Delta}} = \boldsymbol{\tilde{\Delta}}^\text{TPD}$) to quantify the thermal noise transduction across configurations. For $\phi = 0, \pi$, we obtain closed-form expressions for the noise efficiency at the degeneracy, given by
\begin{align}
    \mathrm{N}_\text{th}^\text{TPD}(\phi = 0) &= 4 - \frac{16}{\tilde{\kappa}_c \sqrt{8 - \tilde{\kappa}_c^2}+4},\label{eq:tne_tpd_0} \\
    \mathrm{N}_\text{th}^\text{TPD}(\phi = \pi) &= \frac{2\tilde{\kappa}_c^2 + 8}{\tilde{\kappa}_c^2}.\label{eq:tne_tpd_pi}
\end{align}
Figure~\ref{fig:petermann}(d.iii) depicts Eqs.~\ref{eq:tne_tpd_0}-\ref{eq:tne_tpd_pi} alongside the numerical result for $\phi = \pi/2$. In general, $\mathrm{N}_\text{th}$ tracks the distance between the TPD and the nearest instability transition (Fig.~\ref{fig:petermann}(d.v)), defined as $\text{dist}_\text{Inst}(\phi, \tilde{\kappa}_c) = \min_{\boldsymbol{\tilde{\Delta}} \in \text{Inst}} \lVert \boldsymbol{\tilde{\Delta}}^\text{TPD} - \boldsymbol{\tilde{\Delta}} \rVert_2$. For $\phi = \pi$, $\mathrm{N}_\text{th}$ diverges as $\tilde{\kappa}_c \to 0$ because both $\text{dist}_\text{EP} \to 0$ and $\text{dist}_\text{Inst} \to 0$. By contrast, $\mathrm{N}_\text{th}$ remains finite at $\phi = 0$ for all $\tilde{\kappa}_c \geq 0$ (Appendix \ref{appendix-sec:TNE_away}).

\subsection{\label{subsec:nuisance_scaling}Nuisance Scaling}

Sensors based on degeneracies, like EPs, are notoriously fragile to nuisance fluctuations. Specifically, EP degeneracies are lifted by perturbations in any system parameter, not just the one being sensed. Once the degeneracy is lifted, the square-root splitting immediately vanishes (Appendix \ref{appendix-sec:ep_hypersensitivity}). Thus, infinitesimal fabrication defects, calibration errors, or experimental uncertainty rapidly diminish the enhanced susceptibility. 

Here, we quantify the response of TPDs to these nuisance parameters. We consider a path that crosses the TPD dominated by the nuisance perturbation $\boldsymbol{\tilde{\delta}}$ (with $\boldsymbol{\tilde{\epsilon}} = 0$). For the representative phases $\phi=0$ and $\phi=\pi$, the nuisance perturbation corresponds purely to a frequency $(\tilde{\delta}(\tilde{\Delta}_f))$ and a dissipation $(\tilde{\delta}(\tilde{\Delta}_\kappa))$ shift, respectively. Crossing $\mathrm{Disc} = 0$ with $\boldsymbol{\tilde{\delta}} \neq 0, \boldsymbol{\tilde{\epsilon}} = 0$ confines the system within the single-peak regime, with one peak frequency $\tilde{\nu}_0$. We analyze the Puiseux expansion of the closed form solution for $\tilde{\nu}_0^\text{Root}$ (Appendix \ref{appendix-sec:single_root}) at the TPD, subject to $\boldsymbol{\tilde{\delta}}$, and find 
\begin{align}
    \tilde{\nu}_0^\text{Root}(\phi = 0) &\approx \frac{\lvert \tilde{\kappa}_c^2 - 4\rvert^{1/3}}{2} \text{sgn}(\tilde{\delta}(\tilde{\Delta}_f)) \lvert \tilde{\delta}(\tilde{\Delta}_f) \rvert^{1/3} ,\label{eq:nu_root_phi_0_nuisance} \\
    \tilde{\nu}_0^\text{Root}(\phi = \pi) &\approx \frac{-2^{1/3} \tilde{\kappa}_c^{1/3} (\tilde{\kappa}_c^2 + 4)^{1/6}}{2} \text{sgn}(\tilde{\delta}(\tilde{\Delta}_\kappa))\lvert \tilde{\delta}(\tilde{\Delta}_\kappa) \rvert ^{1/3},\label{eq:nu_root_phi_pi_nuisance}
\end{align}
where the pre-factor of the cube-root term is the nuisance splitting strength ($\tilde{b}_{\mathrm{cbrt}}^{\mathrm{TPD}}$), and $\mathrm{sgn}(\tilde{\delta})=\pm 1$ has the same sign as $\tilde{\delta}$. Figure~\ref{fig:petermann}(d.ii) depicts Eqs.~\ref{eq:nu_root_phi_0_nuisance}–\ref{eq:nu_root_phi_pi_nuisance}. For intermediate values of $\phi$, the behavior is analyzed numerically for $\phi=\pi/2$ in Fig.~\ref{fig:petermann}(d.ii).

\section{\label{sec:ted} Robust Transmission Peak Degeneracies}
\begin{figure*}[htbp!]
\begin{centering}
\includegraphics[width=0.98\textwidth]{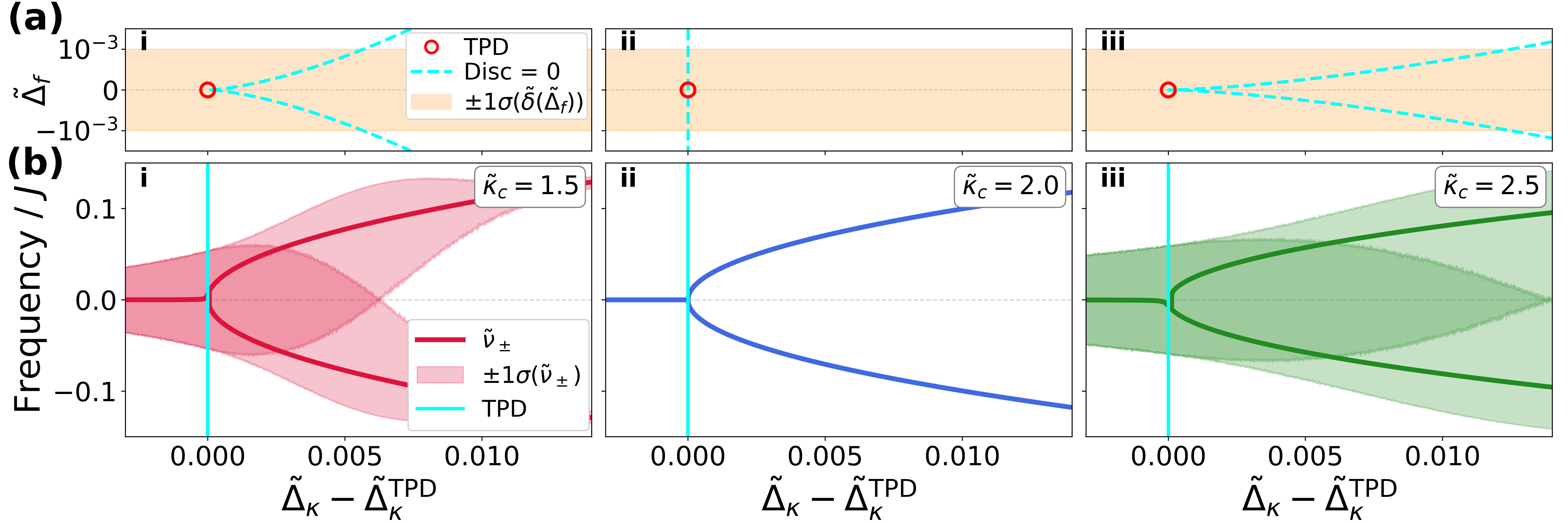}
\end{centering}
\caption{Monte Carlo simulation of nuisance propagation for $\phi = 0$ at three values of $\tilde{\kappa}_c$. (a) Local geometry of the $\mathrm{Disc} = 0$ contour (cyan dashed) near the TPD (red circle). The orange band depicts the standard deviation of a Gaussian-distributed nuisance fluctuation with $\sigma(\tilde{\delta}(\tilde{\Delta}_f)) = 10^{-3}$. In the non-robust configurations (a.i),(a.iii), the $\mathrm{Disc} = 0$ contour forms a cusp catastrophe that intersects the fluctuation band. (a.ii) At the robust TPD  ($\tilde{\kappa}_c = 2$), the contour factorizes and the cusp vanishes. (b) Transduction of $\tilde{\delta}(\tilde{\Delta}_f)$ into transmission peak locations $\tilde{\nu}_\pm$. Solid lines show the mean peak frequencies, while shaded regions indicate $\pm 1\sigma(\tilde{\nu}_\pm)$ propagated from the nuisance fluctuations ($N = 10^4$ samples). Near the TPD, the cusp geometry (b.i),(b.iii) broadens $\sigma(\tilde{\nu}_\pm)$, whereas the robust configuration (b.ii) effectively decouples the peak splitting from $\tilde{\delta}(\tilde{\Delta}_f)$.}
\label{fig:parametric}
\end{figure*}

The cube-root scaling of the nuisance response (Eqs.~\ref{eq:nu_root_phi_0_nuisance}--\ref{eq:nu_root_phi_pi_nuisance}) poses a fundamental vulnerability for TPD-based sensing. Since $\lVert \boldsymbol{\tilde{\delta}} \rVert^{1/3} \gg \lVert \boldsymbol{\tilde{\epsilon}} \rVert^{1/2}$ in the limit of small perturbations, this implies that the system's response to nuisance drift at most TPDs formally exceeds its response to the target signal. This effect is so pronounced that Ref. \cite{xu_single-cavity_2024} exploits it as a sensing mechanism, whereas Ref. \cite{lu_harnessing_2025} treats it as a dominant noise source requiring suppression with closed-loop feedback. 

Equation~\ref{eq:nu_root_phi_0_nuisance} reveals a critical insight into curbing the cube-root nuisance response. For $\phi = 0$, the cube-root nuisance scaling coefficient vanishes exactly at $\tilde{\kappa}_c = 2$. Here, $\tilde{\nu}_0^\text{Root} \approx \tilde{\delta}(\tilde{\Delta}_f)/2$. We term this configuration a robust TPD.

To depict the practical difference between robust and non-robust TPDs, we simulate the propagation of nuisance fluctuations into the transmission spectrum for $\phi = 0$ at three values of $\tilde{\kappa}_c$ (Fig.~\ref{fig:parametric}). We model $\tilde{\delta}(\tilde{\Delta}_f)$ as Gaussian-distributed with $\sigma = 10^{-3}$ (the orange band in Fig.~\ref{fig:parametric}(a) indicates the $\pm 1\sigma$ bounds) and propagate this uncertainty through $N = 10^4$ Monte Carlo samples to obtain the distribution of measured peak frequencies $\tilde{\nu}_\pm$. In the non-robust configurations ($\tilde{\kappa}_c = 1.5$ and $2.5$), the $\mathrm{Disc} = 0$ contour forms a cusp catastrophe \cite{arnold_catastrophe_1992} (Fig.~\ref{fig:parametric}(a.i),(a.iii)), causing significant broadening of the peak distribution, $\sigma(\tilde{\nu}_\pm)$, near the TPD (Fig.~\ref{fig:parametric}(b.i),(b.iii)). In contrast, at the robust configuration ($\tilde{\kappa}_c = 2.0$) $\mathrm{Disc = 0}$ factorizes (Fig.~\ref{fig:parametric}(a.ii)), and $\sigma(\tilde{\nu}_\pm)$ remains suppressed throughout the sweep (Fig.~\ref{fig:parametric}(b.ii)). Further implications of operating a TPD sensor near a cusp catastrophe are explored in Appendix \ref{appendix-sec:cusp_cat}. 

While robust TPDs remove the cubic nuisance scaling, even a linear nuisance response might seem problematic for a degeneracy-based sensor. At EPs, any perturbation, regardless of its scaling, immediately lifts the degeneracy and eliminates the square-root eigenvalue splitting (Appendix~\ref{appendix-sec:ep_hypersensitivity}). Remarkably, TPDs do not share this vulnerability, exhibiting a fundamental advantage of a third-order degeneracy. If a sensing path $\boldsymbol{\tilde{\epsilon}}$ misses the TPD due to $\boldsymbol{\tilde{\delta}} \neq 0$, the path still eventually crosses the $\mathrm{Disc} = 0$ contour at a second-order-degenerate TED. Crucially, crossing a TED preserves the square-root splitting, with
\begin{align}
\tilde{\Delta}_\nu(\boldsymbol{\tilde{\Delta}}^\text{TED}+\boldsymbol{\tilde{\epsilon}}) \approx \tilde{\Delta}_\nu(\boldsymbol{\tilde{\Delta}}^\text{TED}) + \tilde{a}_\text{sqrt}^\text{TED} \sqrt{\lVert \boldsymbol{\tilde{\epsilon}} \rVert}, \label{eq:splitting_expansion}
\end{align}
where $\tilde{a}_\text{sqrt}^\text{TED}$ is the TED splitting strength (derived in Appendix~\ref{appendix-subsec:TED_str}). Thus, the effective sensing target for TPD sensors is not a fragile, isolated point, but an entire surface of TEDs. This intrinsic robustness explains why experimental implementations consistently recover the characteristic square-root response $\tilde{\Delta}_\nu \propto \sqrt{\lVert \boldsymbol{\tilde{\epsilon}} \rVert}$ despite the presence of nuisance fluctuations (Figs. \ref{fig:eps}-\ref{fig:petermann})~\cite{lu_harnessing_2025, kononchuk_exceptional-point-based_2022}.

The robust TPD therefore offers a twofold advantage. First, like all TPDs, it retains the square-root splitting even when nuisance fluctuations displace the sensing path from the exact degeneracy. Second, by removing the cubic dependence of $\tilde \nu_0^\text{Root}$ to nuisance fluctuations, it minimizes the uncertainty propagated into the output spectrum, providing a concrete design target for future TPD-based sensors.

\section{\label{sec:discussion} Discussion}

We have established a unified theoretical framework for TPDs and validated it using a cavity-magnonic platform with full digital control of all system parameters. We systematically investigated TPDs and developed figures of merit for semiclassical sensing, including locations in parameter space, susceptibility to sensing parameters and robustness against nuisance drift, and thermal noise efficiency. Our architecture enables direct sensing of magnetic (through $\tilde{\Delta}_f$) and electric (through $\tilde{\Delta}_\kappa$) fields, with $\phi$ selecting the sensitive parameter. The extensive tunability of our architecture is essential for exploring the full TPD parameter space, but practical sensor deployment will need to replace tunable components with fixed elements chosen according to the design principles established here, particularly the robust TPD condition that suppresses the response to nuisance fluctuations. Alternatively, the active elements can be used to explore nonlinear sensing or generate EP-enhanced magnonic frequency combs \cite{zheng_noise_2025, bai_observation_2024, wang_enhancement_2024}. Beyond sensing, the tunable complex coupling allows for the realization of complex non-Hermitian oscillator lattices that display different topological phases \cite{okugawa_topological_2019, tang_exceptional_2020}, or the study of synthetic photonic materials \cite{owens_quarter-flux_2018, kollar_hyperbolic_2019}, with potential extensions to higher-dimensional EP \cite{zhang_higher-order_2019, hodaei_enhanced_2017} and TPD degeneracies.

\section{\label{sec:acknowledgements} Acknowledgments}

The authors thank Joe Poissant for his support in designing and fabricating the 3D microwave cavities. We thank Hailey A. Mullen, Tian Xia, Sam Sacerdote, Divik Verma, Aidan Curtis, Vincent P. Flynn, Michiel Burgelman, and Lorenza Viola for valuable discussions and feedback on the manuscript. We thank Dr. Mukund Vengalattore for valuable guidance and support of the project. We gratefully acknowledge support from startup funds (Thayer School of Engineering, Dartmouth College), the DARPA Young Faculty Award No.~D23AP00192 (to M.F), and from NSF Grant DGE-2125733 (to support A.S.C).

\section{\label{appendix:a} Appendix A: Derivation of Dynamical Model}

\subsection{Non-reciprocal Dissipative Magnon-Photon Coupling }
We first present a physical motivation of an effective non-Hermitian Hamiltonian for a coupled magnon-photon dimer. The dynamical matrix used for the analysis resembles that of the dissipative magnon-photon coupling Hamiltonian used in cavity magnonics \cite{wang_dissipative_2020}. 
Dissipative magnon-photon coupling was initially understood through a phenomenological electrodynamic framework. In this picture, Ampère's law explains how a microwave current $\mathbf{j}$ generates an AC magnetic field that exerts a driving torque on the magnetization of the YIG sample. In return, dynamic magnetization $\mathbf{m}$ influences the RF current via the Faraday effect. This mutual interaction leads to a form of dissipative coupling, effectively captured by a non-Hermitian term in the description of the system. This term models the backaction of the induced RF current on the magnetization dynamics, acting to impede rather than drive the motion. This framework has laid the foundation for understanding phase nonreciprocity and non-Hermitian dynamics in hybrid magnon–photon systems. In this picture, the following Hamiltonian has been proposed to model the dissipative magnon-photon coupling: \cite{fang_generalized_2017, wang_nonreciprocity_2019, clerk_introduction_2022}

\begin{equation}\label{eq:Hamiltonian}
\hat{H} =  \omega_1 \hat{a}_1^\dagger \hat{a}_1 +  \omega_2 \hat{a}_2^\dagger \hat{a}_2 + J \left( \hat{a}_1^\dagger \hat{a}_2 + e^{i\phi} \hat{a}_2^\dagger \hat{a}_1 \right),
\end{equation}
where $\omega_{1}$ and $\omega_{2}$ represent cavity and the magnon mode frequencies respectively. Notice that for $\phi=0$ and $\phi=\pi$ the coupling becomes coherent and for other phases it is both dissipative and coherent. In the subsequent section, we adapt this Hamiltonian and obtain equations of motion in the semiclassical limit.
Non-reciprocal hopping in Eq.~\ref{eq:Hamiltonian} can emerge due to dissipative effects mediated by a shared environment. It can be thought of as a non-Hermitian beam-splitter. For completeness, we would like to briefly summarize a quantum model and present an argument for phase non-reciprocity so as to motivate the dynamical matrix used in this work. The microscopic derivation of non-reciprocal magnon-photon interactions using traveling wave amplifiers has been presented here \cite{yao_microscopic_2019}. 

We consider two coupled modes, both coherently and dissipatively through a shared reservoir \cite{metelmann_nonreciprocal_2015}. The coherent interaction is described by the Hamiltonian
\[
\hat{H}_{\text{coh}} = \hbar J \left( \hat{a}^\dagger \hat{b} + \hat{b}^\dagger \hat{a} \right),
\]
where \( J \) is the coherent coupling strength.

Dissipative coupling is introduced through a shared reservoir, characterized by the jump operator
\[
\hat{o} = \hat{a} + r e^{i\theta} \hat{b},
\]
where \( r \) is a real amplitude ratio and \( \theta \) is the relative phase. This yields a non-Hermitian contribution to the effective Hamiltonian,
\[
\hat{H}_{\text{diss}} =  -\frac{i\gamma}{2} \hat{o}^\dagger \hat{o} 
=  -\frac{i\gamma}{2} \left( \hat{a}^\dagger + r e^{-i\theta} \hat{b}^\dagger \right) \left( \hat{a} + r e^{i\theta} \hat{b} \right).
\]
Expanding this equation gives the diagonal and the off-diagonal terms, and the dissipative part becomes
\begin{equation}
\hat{H}_{\text{diss}}=-\frac{i\gamma}{2}\left(a^{\dagger}a+r^2b^{\dagger}b\right)-\frac{i\gamma r}{2}
\left(e^{i\theta}ab^{\dagger}+e^{-i\theta}b^{\dagger}a\right)
\end{equation}

Combining the coherent and dissipative interactions, the off-diagonal part of the Hamiltonian becomes
\begin{equation}
 H_{o}=\left(J - i \frac{\gamma r}{2} e^{-i\theta}\right)a^{\dagger}b+\left(J - i \frac{\gamma r}{2} e^{i\theta}\right)b^{\dagger}a
\end{equation}

To achieve unidirectional coupling, we set \( \theta = \frac{\pi}{2} \), so that \( e^{\pm i\theta} = \pm i \), and choose \( J = \frac{\gamma r}{2} \). Substituting these values, we find
\[
\begin{aligned}
\hat{a}^\dagger \hat{b} &: \quad  J - i \frac{\gamma r}{2} (-i) =  J + \frac{\gamma r}{2} = \gamma r, \\
\hat{b}^\dagger \hat{a} &: \quad  J - i \frac{ \gamma r}{2} (i) =  J - \frac{\gamma r}{2} = 0.
\end{aligned}
\]
This configuration yields perfect non-reciprocity: excitations transfer from mode \( a \) to mode \( b \), but not in the reverse direction.

In the general case, tuning the amplitude \( r \) and phase \( \theta \) of the dissipative off-diagonal coupling suggests the following form of an effective non-Hermitian Hamiltonian 
\[
\hat{H}_{\text{eff}} = 
 \begin{pmatrix}
\omega_a - i \kappa_a & J \\
J e^{i\phi} & \omega_b - i \kappa_b
\end{pmatrix},
\]
where \( J \) is real, and the complex phase \( \phi \) introduces directional asymmetry between the two modes and 
$\kappa_a=\frac{\gamma}{2}$ and $\kappa_{b}=\frac{\gamma r^{2}}{2}$. This structure forms the basis for engineering nonreciprocal dynamics in open quantum systems.
In the semi-classical limit it can be thought of as a dynamical non-Hermitian matrix realized in our room temperature set-up.
This essentially describes coupled oscillators with broken reciprocity. 

\subsection{Classically Coupled Oscillators with Broken Reciprocity}
The form of the dynamical matrix used in this work can also be obtained purely from a classical model of interacting harmonic oscillators with broken time reversal symmetry. 
We consider two coupled classical oscillators \( x_1(t) \) and \( x_2(t) \) with damping and a non-reciprocal phase in the coupling. The equations of motion are:

\begin{align}
\ddot{x}_1 + \gamma_1 \dot{x}_1 + \omega_1^2 x_1 + J^2 x_2 &= 0, \\
\ddot{x}_2 + \gamma_2 \dot{x}_2 + \omega_2^2 x_2 + J^2 e^{i \phi} x_1 &= 0,
\end{align}

where
\begin{itemize}
  \item \( \omega_1, \omega_2 \) are the natural frequencies of the oscillators,
  \item \( \gamma_1, \gamma_2 \) are the damping coefficients,
  \item \( K \) is the real-valued coupling strength,
  \item \( \phi \) is a non-reciprocal phase (appearing only in one direction of coupling).
\end{itemize}
This can be written in matrix form as follows.
Define \( \mathbf{x}(t) = \begin{pmatrix} x_1(t) \\ x_2(t) \end{pmatrix} \), then the equations of motion can be written as:

\begin{equation}
\ddot{\mathbf{x}} + \Gamma \dot{\mathbf{x}} + D \mathbf{x} = 0,
\end{equation}
where
\[
\Gamma = 
\begin{pmatrix}
\gamma_1 & 0 \\
0 & \gamma_2
\end{pmatrix}, \qquad
D =
\begin{pmatrix}
\omega_1^2 & J^2 \\
J^2 e^{i \phi} & \omega_2^2
\end{pmatrix}.
\]
A similar model has been used in the context of dissipative magnon-photon interaction within the radiation damping picture \cite{wang_dissipative_2020}

\subsection{Theory for the Linear Model: Equations of Motion}
The equations of motion presented in the draft can be interpreted as the semiclassical limit of an effective non-Hermitian Hamiltonian. This effective Hamiltonian incorporates diagonal self-energy terms to model losses, while the off-diagonal self-energy terms arise from dissipative, phase-dependent hopping. More importantly, the off-diagonal terms can have a nonreciprocal nature. Such a dissipative coupling can be engineered using structured reservoirs such as waveguides and by introducing synthetic gauge fields. 
We consider a system of coupled oscillators. The effective Hamiltonian for this system in the rotating frame (rotating with drive frequency) can be written as ($\hbar=1$):

\begin{align}
H_{\text{eff}} &= (\omega_c - \omega_d) a_c^\dagger a_c + (\omega_y - \omega_d) a_y^\dagger a_y \\ \nonumber
&\quad - i \frac{\kappa_{c}}{2} a_c^\dagger a_c - i\frac{\kappa_{y}}{2} a_y^\dagger a_y \\ \nonumber
&\quad + J\left( a_c^\dagger a_y + e^{-i\phi}a_y^\dagger a_c \right) \\ \nonumber
&\quad + f_c (a_c^\dagger + a_c)+f_y (a_y^\dagger + a_y)
\end{align}
 We consider two non-interacting cavities given by the following Hamiltonian:
\begin{equation}
    H_{0}=(\omega_{c}-\omega_{d}) a_{c}^{\dagger}a_{c}+(\omega_{y}-\omega_{d}) a_{y}^{\dagger}a_{y}+u_c (a_c^\dagger + a_c)+ u_y (a_y^\dagger + a_y)
\end{equation}
The effective non-Hermitian Hamiltonian for cavities coupled to a bath can be written as 
\begin{equation}\label{eq:dissipator}
H_{\text{eff}}=H_{0}-\frac{i}{2}{\bf L}^{\dagger}{\bf \Gamma}\bf{L},
\end{equation}
where $\mathbf{L} = \begin{pmatrix} \hat{a}_1 \\ \hat{a}_2 \end{pmatrix}.
$
The effective Hamiltonian in Eq.~\ref{eq:dissipator} can be written in the following form 
\begin{equation}
\hat{H}_{\mathrm{eff}} = H_{0} 
- \frac{i}{2}
\begin{pmatrix}
\hat{a}_c^\dagger & \hat{a}_y^\dagger
\end{pmatrix}
\begin{pmatrix}
\kappa_{c} & 2iJ \\
2iJ e^{-i\phi} & \kappa_{y}
\end{pmatrix}
\begin{pmatrix}
\hat{a}_c \\
\hat{a}_y
\end{pmatrix}
\end{equation}

We can see that the diagonal terms give us a loss that makes oscillator frequencies complex, and the off-diagonal terms give us the desired nonreciprocal coupling.
Any non-Hermitian Hamiltonian can be written as a sum of a Hermitian and an anti-Hermitian matrix \cite{brody_coherent_2010}. To do so, consider a non-Hermitian Hamiltonian operator \( \hat{H} \neq \hat{H}^\dagger \), which determines the time evolution of an arbitrary quantum observable \( \hat{\chi} \) via the commutator:
\begin{equation}
    i \frac{d\hat{\chi}}{dt} = [\hat{\chi}, \hat{H}].
\end{equation}
This commutator can be expressed in matrix form as:
\begin{equation}
    i \frac{d\hat{\chi}}{dt} = 
    \begin{bmatrix}
        \hat{\chi} & \hat{H}
    \end{bmatrix}
    \
    \begin{bmatrix}
        0 & 1 \\ 
        -1 & 0
    \end{bmatrix}
    \
    \begin{bmatrix}
        \hat{\chi} \\ 
        \hat{H}
    \end{bmatrix}.
\end{equation}
Here, the symplectic matrix is defined as:
\begin{equation}
    \Omega = 
    \begin{bmatrix}
        0 & 1 \\ 
        -1 & 0
    \end{bmatrix}.
\end{equation}
For compact notation, we define the column vector:
\begin{equation}
    \chi_{\hat{H}} = 
    \begin{bmatrix}
        \hat{\chi} \\ 
        \hat{H}
    \end{bmatrix}.
\end{equation}
The above equation can be written in the form 
\begin{equation}
i \frac{d\chi}{dt}  = \chi^T H \, \Omega \, \chi \\
= [\chi(t), H]_\Omega
\end{equation}
 The above equation demonstrates the equivalence of this matrix structure concerning the commutator, establishing a Lie bracket and Lie algebra framework in quantum mechanics. By emphasizing the role of the matrix $\Omega$, it suggests that alternative algebras and brackets can be constructed. This motivates the decomposition of any generic non-Hermitian Hamiltonian into Hermitian and anti-Hermitian parts.
The generic non-Hermitian Hamiltonian can always be decomposed into its Hermitian and anti-Hermitian components:
\begin{equation}
    \hat{H} = \hat{H}_+ + \hat{H}_-,
\end{equation}
where
\begin{equation}
    \hat{H}_+ = \frac{\hat{H} + \hat{H}^\dagger}{2}, \quad 
    \hat{H}_- = \frac{\hat{H} - \hat{H}^\dagger}{2i}.
\end{equation}

The commutator with the non-Hermitian Hamiltonian splits into two parts
\begin{equation}
    [\hat{\chi}, \hat{H}] = [\hat{\chi}, \hat{H}_+] + [\hat{\chi}, \hat{H}_-].
\end{equation}

From these equations, the equation of motion for the density matrix emerges \cite{brody_coherent_2010}, incorporating both the commutator and the anti-commutator:
\begin{equation}
    \frac{d\hat{\rho}(t)}{dt}  = -i[\hat{H}_+, \hat{\rho}(t)] + \{i\hat{H}_-, \hat{\rho}(t)\}_+,
\end{equation}
where \( \{\cdot, \cdot\}_+ \) denotes the anticommutator.
The above equation resembles the Lindblad master equation, but notice that the dissipation function is given by an anti-commutator term that contains the anti-Hermitian part of the Hamiltonian \cite{brody_coherent_2010}.
In the classical limit, the above equations of motion can be expressed as below:
\begin{equation}
    \begin{bmatrix}
        \dot{q} \\ 
        \dot{p}
    \end{bmatrix}
    = \bm{\Omega}^{-1} \nabla H - \bm{G}^{-1} \nabla {{\Sigma}},
\end{equation}
where the dynamical equation is described in terms of two mathematical objects and their associated flows. 

Here, \( \bm{\Omega} \) is the symplectic unit matrix, and
\( \bm{G} \) is the phase space metric that describes the Euclidean geometry of phase space. The second flow is the canonical metric gradient flow, which arises from the imaginary part, \( \Sigma \), of the Hamiltonian function. This flow represents the dissipative aspect of the dynamics, where energy gradients drive the evolution, and it is crucial for systems with non-conservative forces or open quantum systems.
In the semi-classical limit, the creation and annihilation operators can be replaced by amplitudes, and the corresponding coupled equations of motion have the following form.

\begin{widetext}
\begin{equation}
\label{eq:alphas}
\frac{d}{dt}
\begin{pmatrix}
\alpha_c \\
\alpha_y
\end{pmatrix}
=
\begin{pmatrix}
-i(\omega_c - \omega_d - i\frac{\kappa_c}{2}) & -iJ \\
-iJ e^{i\phi} & -i(\omega_y - \omega_d - i\frac{\kappa_y}{2})
\end{pmatrix}
\begin{pmatrix}
\alpha_c \\
\alpha_y
\end{pmatrix}
+
\begin{pmatrix}
u_c \\
u_y
\end{pmatrix}.
\end{equation}
\end{widetext}

The two-by-two matrix in Eq.~\ref{eq:alphas} is the dynamical matrix used for further analysis in the linear regime.
We can also derive the same equations of motion by directly applying the Heisenberg equations of motion to an effective non-Hermitian Hamiltonian. The above dynamical matrix can also be obtained by introducing an intermediate third-mode approach, as shown in Ref. \cite{rohn_classical_2023}.

To make Eq.~\ref{eq:alphas} more amenable to an engineering context, we make the following substitutions. First, we replace angular frequencies such as $\omega_c$ with $f_c = \omega_c/(2\pi)$. Next, we nondimensionalize the model by dividing all parameters by $J$, adding a tilde on top of all parameters to designate nondimensionalized parameters. Then, we remove the drive frequency $\tilde f_d$ from the dynamical matrix. Finally, we make the following substitutions, which will end up simplifying further analysis:
\begin{align}
    \tilde \Delta_f &\equiv \tilde f_c - \tilde f_y \\ 
    \tilde \Delta_\kappa &\equiv (\tilde \kappa_c - \tilde \kappa_y)/2
\end{align}
\subsection{\label{appendix-sec:petermann}Petermann Factor Analysis}
Since the radicand of Eq.~\ref{eq:deltalambda_tilde} is generally complex, the square-root is computed using the principal value, which we can define in polar form as $\tilde \Delta_\lambda = \sqrt{|\tilde \Delta_\lambda|^2}\, e^{\mathrm{i}\arg(\tilde \Delta_\lambda^2)/2}$, with the argument $\arg(\tilde \Delta_\lambda^2) \in (-\pi, \pi]$. This choice ensures continuity of $\tilde \Delta_\lambda$ in parameter space except at branch points along the zero contour $\tilde \Delta_\lambda = 0$. This zero contour is a curve along which the two eigenvalues are degenerate in real and imaginary components. To determine whether these degeneracies are ordinary crossings or EPs, we examine the eigenvector coalescence, quantified by the Petermann noise factor. 

The Petermann factor is defined as $K_{j'j} = \langle L_{j'}|L_j\rangle \langle R_j|R_{j'}\rangle$, where $\langle L_j|$ and $|R_j\rangle$ denote the left and right eigenvectors of $\bm{\tilde{A}}$ \cite{siegman_excess_1989, ashida_non-hermitian_2020}. For convenience, we use the mean diagonal Petermann factor $\text{PF} = \frac{1}{2}(K_{11} + K_{22})$, whose deviation from unity measures eigenvector non-orthogonality and diverges at EPs \cite{zheng_mathcalpt_2010}. While originally formulated to describe excess spontaneous emission in unstable lasers \cite{siegman_excess_1989}, the Petermann factor now serves more generally as a predictor of noise enhancement in non-Hermitian systems \cite{wang_petermann-factor_2020, ghosh_suppression_2024}. To compute an analytical equation for the Petermann factor, we closely follow an example computation from Ref. \cite{ashida_non-hermitian_2020}. First, we compute the eigenprojectors $P_\pm$ using 
\begin{equation}
    P_\pm = -\oint_{C_\pm} \frac{dz}{2\pi i} R(z) = \text{Res}_{z=\lambda_\pm} R(z)
\end{equation}
Here $R(z) = (\bm{\tilde{A}} - zI)^{-1}$. To compute $P_\pm$, we use symbolic residue calculation in a computer algebra system. The spectral projector \( P_\pm \) is given by the residue at \( z = \tilde \lambda_\pm \), which corresponds to the coefficient of the \( (z - \tilde \lambda_\pm)^{-1} \) term. Multiplying the truncated series by \( (z - \tilde \lambda_\pm) \) cancels the pole, leaving just the numerator of the singular term, the residue itself. We can then compute the Petermann factor using
\begin{equation}
    \text{PF} = \frac{1}{2} (||P_+||_2^2 + ||P_-||_2^2),
\end{equation}
where $||M||_2 = \sqrt{\text{Tr}({M^\dagger M})}$ refers to the Frobenius (Hilbert-Schmidt) matrix norm. This yields the final equation for the Petermann factor, given in the Main Text as
\begin{equation}\label{eq:supp_peterman}
\text{PF} = \frac{\tilde \Delta_f^2 + \tilde \Delta_\kappa^2 + |\tilde \Delta_\lambda|^2 + 4}{2|\tilde \Delta_\lambda|^2}.
\end{equation}

\section{\label{appendix:b} Appendix B: Complete Parameterization of Surface of Transmission Peak Degeneracies}

Here, we provide the full parameterization of the surface of TPDs surveyed in the Main Text. For $\phi = 0$, there are two TPDs along the $\tilde \Delta_\kappa$ axis for $\tilde \kappa_c \leq 2\sqrt{2}$, located at $\frac{\tilde \kappa_c}{2} \pm \frac{\sqrt{8 - \tilde \kappa_c^2}}{2}$. When $\tilde \kappa_c \geq 2$, four more rogue TPDs appear, with locations $(\tilde \Delta_\kappa, \tilde \Delta_f) = (0, \pm \sqrt{\tilde \kappa_c^2 -4})$, and $(\tilde \Delta_\kappa, \tilde \Delta_f) = (\tilde \kappa_c, \pm \sqrt{\tilde \kappa_c^2 -4})$. For $\phi = \pi$, there are generally four TPDs, located at $(\tilde \Delta_\kappa, \tilde \Delta_f) = (0, \pm \sqrt{\tilde \kappa_c^2 +4})$, and rogue TPDs located at $(\tilde \Delta_\kappa, \tilde \Delta_f) = (\tilde \kappa_c, \pm \sqrt{\tilde \kappa_c^2 +4})$. For all other values of $\phi$, the locations of the TPDs require solving the following expression given by
{\small
\begin{align}
    x &\equiv \text{roots}\!\left(
        2x^4 - 2\tilde \kappa_c x^3
        - x^2(4\cos(\phi) - \tilde \kappa_c^2)
        + 4\cos^2(\phi) - 4
    \right), \\
    \tilde \Delta_f &= - \frac{-\tilde \kappa_c^2 x + 2 \tilde \kappa_c x^2 - 2x^3 + 4 \cos(\phi)x}{2\sin(\phi)}, \\
    \tilde \Delta_\kappa &= x.
\end{align}
}
where the $\text{roots}$ operation refers to finding only the roots of the quartic equation without an imaginary component such that $x\in \mathbb{R}$. Furthermore, there are two rogue TPDs at $(\tilde \Delta_\kappa, \tilde \Delta_f) = (\tilde \kappa_c, \pm \sqrt{\tilde \kappa_c^2 - 4\cos(\phi)})$.

\section{\label{appendix:c} Appendix C: Additional Results on Exceptional Points and Transmission Peak Degeneracies}

\subsection{\label{appendix-sec:ep_hypersensitivity}Exceptional Point Nuisance Response}

Unlike TPDs, which can be engineered for robustness against nuisance parameters, EPs exhibit divergent susceptibility in all parameter directions simultaneously. This fundamentally limits EP-based sensing in the presence of nuisance fluctuations.

We analyze the eigenvalue splitting (Eq.~\ref{eq:deltalambda_tilde}) near an EP by introducing perturbations in both the sensing and nuisance directions. For $\phi = 0$, we consider the EP at $(\tilde \Delta_\kappa^\text{EP}, \tilde \Delta_f^\text{EP}) = (+2, 0)$. Following the optimal sensing path $\tilde q = 0$, the sensing perturbation is $\tilde \epsilon(\tilde \Delta_\kappa)$, while $\tilde \delta(\tilde \Delta_f)$ acts as a nuisance. Substituting $\tilde \Delta_\kappa = 2 - \tilde \epsilon$ and $\tilde \Delta_f = \tilde \delta$ into Eq.~\ref{eq:deltalambda_tilde} yields
\begin{align}
    \operatorname{Im}(\Delta_\lambda) = \operatorname{Im}\left( \sqrt{(\tilde \epsilon - 2)^2 - \tilde \delta^2 - 4 - 2\tilde \delta(\tilde \epsilon - 2)i} \right).
\end{align}
In the case of perfect alignment ($\tilde \delta = 0$), the splitting reduces to $\operatorname{Im}(\tilde \Delta_\lambda) \approx 2\sqrt{\tilde \epsilon}$, yielding a susceptibility
\begin{align}
    \tilde \chi_{\tilde \epsilon(\tilde \Delta_\kappa)}^\text{EP} = \frac{\partial}{\partial \tilde \epsilon}\operatorname{Im}(\tilde \Delta_\lambda) \bigg|_{\tilde \delta = 0} \propto \frac{\tilde{a}_{\text{sqrt}}^\text{EP}}{2\sqrt{\tilde \epsilon}} \to \infty \quad \text{as } \tilde \epsilon \to 0,
\end{align}
with sensing splitting strength $\tilde{a}_{\text{sqrt}}^\text{EP}=2$. By contrast, in the case of nuisance-only perturbations ($\tilde \epsilon = 0$), the splitting becomes $\operatorname{Im}(\tilde \Delta_\lambda) \approx \sqrt{2}\sqrt{\tilde \delta}$, with susceptibility
\begin{align}
    \tilde \chi_{\tilde \delta(\tilde \Delta_f)}^\text{EP} = \frac{\partial}{\partial \tilde \delta}\operatorname{Im}(\tilde \Delta_\lambda) \bigg|_{\tilde \epsilon = 0} \propto \frac{\tilde{b}_{\text{sqrt}}^\text{EP}}{2\sqrt{\tilde \delta}} \to \infty \quad \text{as } \tilde \delta \to 0,
\end{align}
with nuisance splitting strength $\tilde{b}_{\text{sqrt}}^\text{EP}=\sqrt{2}$. Finally, with $\tilde \delta \neq 0, \tilde \epsilon \neq 0$, evaluating the sensing susceptibility at finite $\tilde \delta$ yields
\begin{align}
    \lim_{\tilde \epsilon \to 0^+} \tilde \chi_{\tilde \epsilon(\tilde \Delta_\kappa)}^\text{EP} \bigg|_{\tilde \delta \neq 0} \propto \operatorname{Im}\left( \frac{4 + 2\tilde \delta \, i}{\sqrt{-\tilde \delta^2 + 4\tilde \delta \, i}} \right) < \infty.
\end{align}
Thus, any nonzero nuisance perturbation $\tilde \delta$ regularizes the divergent susceptibility, reducing it to a finite value. The same structure holds for $\phi = \pi$ with the roles of $\tilde \Delta_\kappa$ and $\tilde \Delta_f$ exchanged, and for general $\phi$ along the hyperbolic sensing path. Thus, for EP-based sensors, the square-root response that yields enhanced susceptibility is nearly equally sensitive to calibration errors, fabrication tolerances, and environmental drift as it is to the target signal.

\subsection{\label{appendix-sec:geometric_interp}Geometric Interpretation of Peaks and Imaginary Eigenvalues}

Here, we include an additional perspective on the interpretation of peak locations, eigenvalues, EPs, and TPDs. There are special conditions for $\tilde \kappa_c$ such that EPs and TPDs coalesce into an EP-TPD pair. In Fig.~\ref{supp_fig:combined_drive_vector_geometric}, we show that these conditions occur for $\phi = 0$ at $\tilde \kappa_c = 2$ (Fig.~\ref{supp_fig:combined_drive_vector_geometric}(a.iii)), and for $\phi = \pi$ at $\tilde \kappa_c = 0$ (Fig.~\ref{supp_fig:combined_drive_vector_geometric}(b.i)). However, for all other $\tilde \kappa_c$, the locations of TPDs, EPs, $\tilde \nu_\pm$, and $|\text{Im}(\tilde \lambda_\pm)|$ are distinct. In particular, for $\phi = 0$, $\tilde \nu_\pm$ form ellipses whose semi-major ($\tilde \Delta_\kappa$) and semi-minor (frequency) axes scale with $\tilde \kappa_c$, while the ellipses formed by $|\text{Im}(\tilde \lambda_\pm)|$ stay constant with $\tilde \kappa_c$, providing a bound on $\tilde \nu_\pm$. For $\phi = \pi$, $\tilde \nu_\pm$ and $|\text{Im}(\tilde \lambda_\pm)|$ form hyperbolae, where the $\tilde \nu_\pm$ hyperbolae vertices drift outwards with $\tilde \kappa_c$. 

\subsection{\label{appendix-sec:drive_readout_vector}Effect of Drive and Readout Vector on Peak Locations}

\begin{figure*}
\begin{centering}
\includegraphics[width=0.98\textwidth]{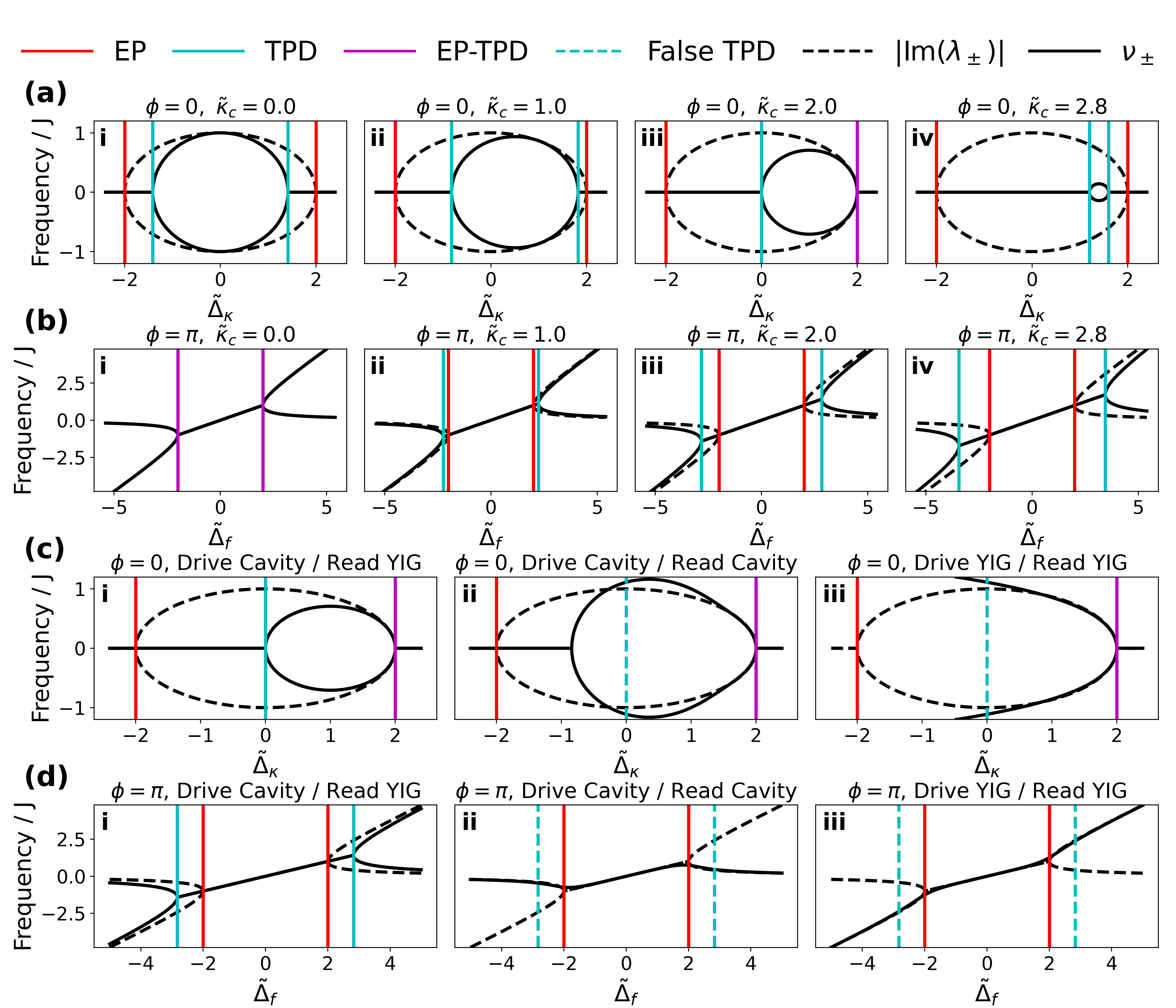}
\end{centering}
\caption{(a)(b) Geometric interpretation of the $\phi = 0$ and $\phi = \pi$ configurations. The imaginary eigenvalues, $|\text{Im}(\tilde \lambda_\pm)|$ do not change with $\tilde \kappa_c$, while the peak locations, $\tilde \nu_\pm$, do. Furthermore, the peak locations do not line up with the imaginary eigenvalues, but remain bounded by them. (c)(d) Simulations of $\tilde \nu_\pm$ for $\tilde \kappa_c =2$. \textbf{a(d)} depicts drive cavity, readout YIG $\bm{B} = [1,0]^{\mathrm T}, \bm{C} = [0, 1]$ for $\phi = 0$ ($\phi = \pi$), where theory presented in the Main Text holds. \textbf{b(e)} depicts drive/readout cavity configuration $\bm{B} = [1,0]^{\mathrm T}, \bm{C} = [1,0]$ for $\phi = 0$ ($\phi = \pi$). \textbf{c(f)} depicts drive/readout YIG configuration $\bm{B} = [0, 1]^{\mathrm T}, \bm{C} = [0, 1]$ for $\phi = 0$ ($\phi = \pi$). Dashed cyan lines depict mismatch between Main Text $\boldsymbol{\tilde \Delta}^\text{TPD}$ theory and alternate drive vectors.}
\label{supp_fig:combined_drive_vector_geometric}
\end{figure*}

The Main Text considers only the configuration where the cavity is driven ($\bm{B} = [1,0]^{\mathrm T}$) and the YIG is read out ($\bm{C} = [0, 1]$), which admits exact solutions for the peak roots $\tilde{\nu}_\pm^\text{Root}$. Other input-output configurations yield qualitatively different behavior (Fig.~\ref{supp_fig:combined_drive_vector_geometric}). For driving and reading out the cavity ($\bm{B} = [1,0]^{\mathrm T}$, $\bm{C} = [1, 0]$), $\phi = 0$ admits two TPDs (one not predicted by the Main Text theory (Fig.~\ref{supp_fig:combined_drive_vector_geometric}(c.ii))) while $\phi = \pi$ admits none (Fig.~\ref{supp_fig:combined_drive_vector_geometric}(d.ii)). For driving and reading out the YIG ($\bm{B} = [0, 1]^{\mathrm T}$, $\bm{C} = [0, 1]$), $\phi = 0$ admits one TPD (Fig.~\ref{supp_fig:combined_drive_vector_geometric}(c.iii)) and $\phi = \pi$ admits none (Fig.~\ref{supp_fig:combined_drive_vector_geometric}(d.iii)).

\subsection{\label{appendix-sec:single_root}Single Root Closed Form Solution}

When $\text{Disc} < 0$, Eqs.~\ref{eq:theta}-\ref{eq:nu_minus} are invalid, as there is only one real root, $\tilde \nu_0^\text{Root}$, given by 
\begingroup
\small
\begin{equation}\label{eq:hyperbolic_root}
    \tilde \nu_0^\text{Root} = 
    \begin{cases}
        -2 \dfrac{|\tilde q|}{\tilde q} \sqrt{\dfrac{-\tilde p}{3}} 
        \cosh\!\left[
            \dfrac{1}{3} \, \mathrm{arcosh} \left(
                \dfrac{-3|\tilde q|}{2\tilde p} \sqrt{\dfrac{-3}{\tilde p}}
            \right)
        \right],
        & \text{if }  \tilde p < 0, \\[1.5ex]
        -2 \sqrt{\dfrac{\tilde p}{3}} 
        \sinh\!\left[
            \dfrac{1}{3} \, \mathrm{arsinh} \left(
                \dfrac{3\tilde q}{2\tilde p} \sqrt{\dfrac{3}{\tilde p}}
            \right)
        \right],
        & \text{if } \tilde p > 0.
    \end{cases}
\end{equation}
\endgroup
Equation~\ref{eq:hyperbolic_root} is analyzed using a Puiseux expansion at the TPD in the $\boldsymbol{\tilde \delta}$ direction to obtain the nuisance scaling (Eqs.~\ref{eq:nu_root_phi_0_nuisance}-\ref{eq:nu_root_phi_pi_nuisance}). 

\subsection{\label{appendix-sec:TNE_away}Thermal Noise Efficiency Away From Transmission Peak Degeneracies}

The results in Sec \ref{subsec:additive_noise} imply that the thermal noise efficiency ($\mathrm{N}_\text{th}$) is closely related to the Petermann noise factor. However, while the expressions for the two noise factors are similar at $\boldsymbol{\tilde \Delta}^\text{TPD}$ coordinates, it is not a general relationship. In general, $\mathrm{N}_{\text{th}}$ diverges at instability transitions, while the Petermann factor diverges at EPs. Figure~\ref{supp_fig:tne} depicts $\mathrm{N}_{\text{th}}$ calculated on $\boldsymbol{\tilde \Delta}$-space, with exactly the same parameters as the Petermann factor landscape in Fig.~\ref{fig:theory}, depicting a significant difference in the behavior of the $\mathrm{N}_{\text{th}}$ for parameters away from $\boldsymbol{\tilde \Delta}^\text{TPD}$. 

\begin{figure}
\centering
\includegraphics[width=0.49\textwidth]{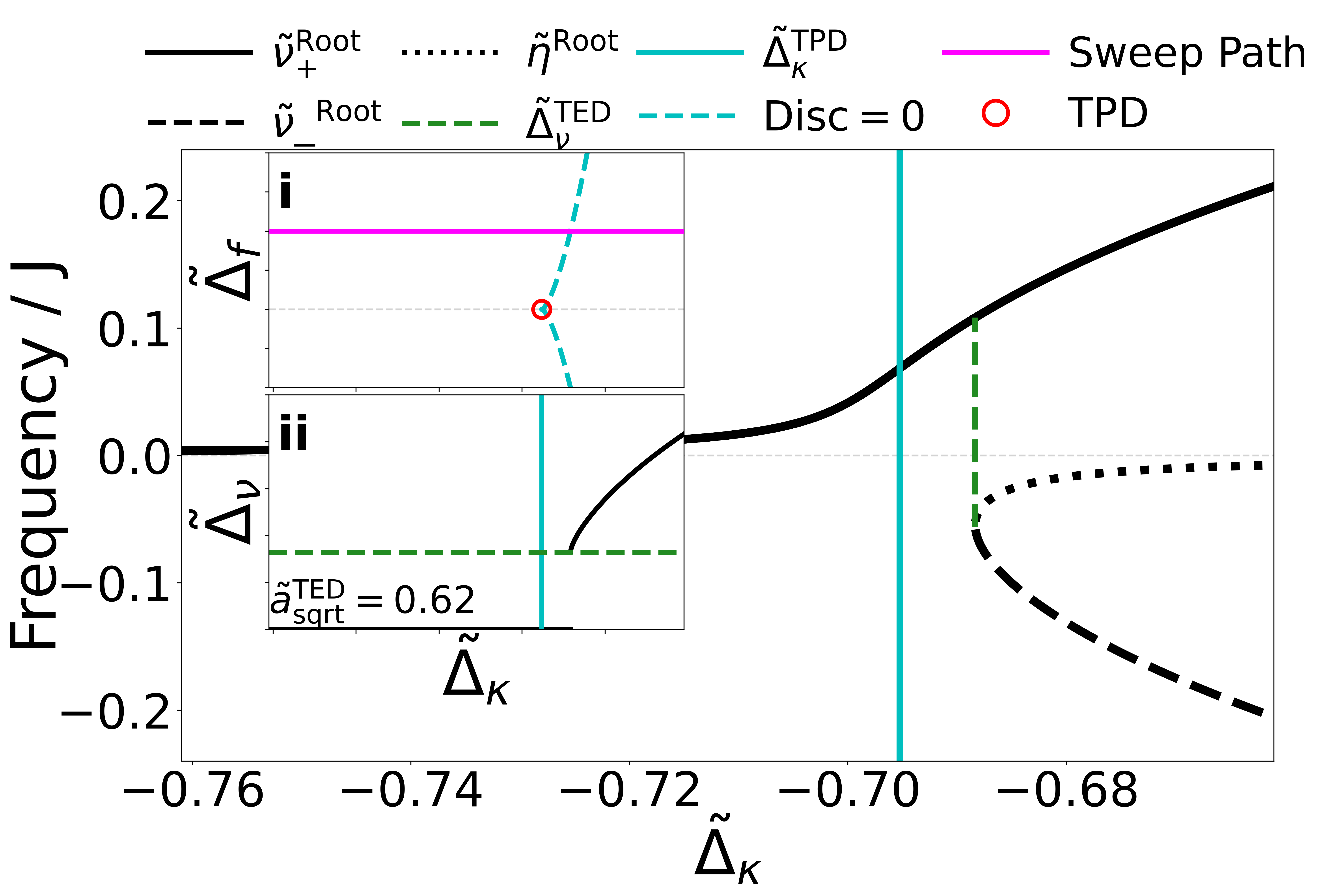}
\caption{Peak splitting at a TED. (a.i) local scaling of the cusp catastrophe near the TPD. A sweep path (magenta line) with $\boldsymbol{\tilde \delta} = 10^{-3}$ intersects $\mathrm{Disc = 0}$ at a TED, retaining square-root splitting with $\tilde{a}_\text{sqrt}^\text{TED} > 0$ (a.ii). At the TED, the residual splitting $\tilde \Delta_\nu^\text{TED} \neq 0$.}
\label{fig:TED_single_panel}
\end{figure}

\section{\label{appendix-sec:cusp}Transmission Extrema Degeneracies}

TEDs occur along the $\mathrm{Disc} = 0$ contour where two roots of Eq.~\ref{eq:cubic} coalesce. While TPDs are third-order degeneracies where all three roots merge ($\tilde{\nu}_+ = \tilde{\nu}_- = \tilde{\eta}$), TEDs are second-order degeneracies where one transmission peak coalesces with the local minimum ($\tilde{\nu}_\pm = \tilde{\eta}$), leaving the other peak isolated. Crucially, a sensing path displaced from the TPD by nuisance fluctuations will still eventually cross the $\mathrm{Disc} = 0$ contour at a TED, retaining square-root splitting. 

\subsection{\label{appendix-sec:vieta}Residual Splitting at a TED}

\begin{figure}
\centering
\includegraphics[width=0.48\textwidth]
{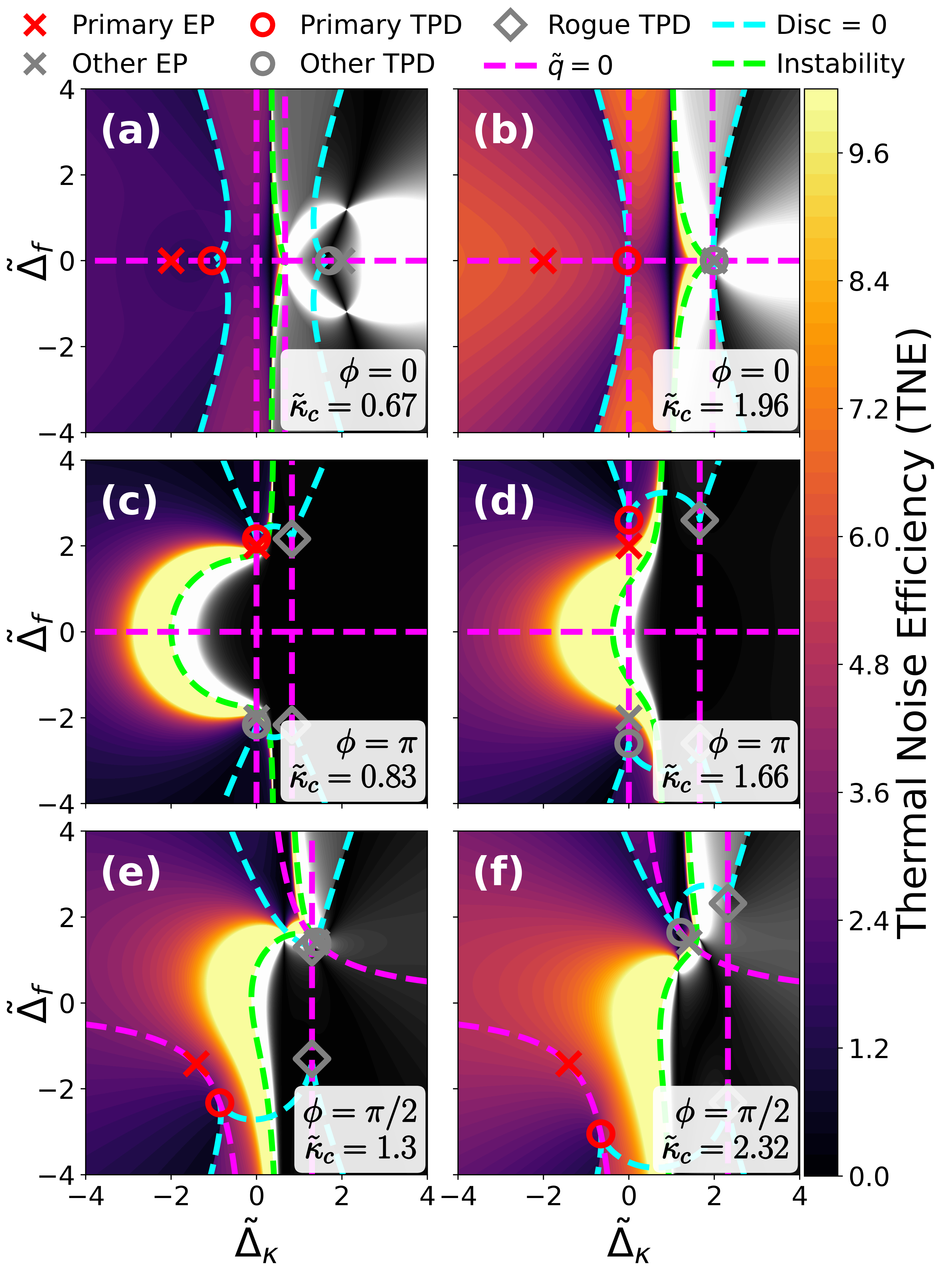}
\caption{Theory for $\phi = 0$ (a),(b), $\phi = \pi$ (c),(d), and $\phi = \pi/2$ (e),(f) overlaid atop $\mathrm{N}_\text{th}$ (clipped to 90th percentile), using the exact parameters as Fig.~\ref{fig:theory} to depict the contrast between the Petermann factor and $\mathrm{N}_\text{th}$.}
\label{supp_fig:tne}
\end{figure}

At a TPD, $\tilde \Delta_\nu = 0$. This is not the case at TEDs. Instead, $\tilde{\Delta}_\nu(\boldsymbol{\tilde{\Delta}}^\text{TED}) \neq 0$. We denote this residual peak splitting as $\tilde \Delta_\nu^\text{TED}$ for brevity. Here, we derive the residual peak splitting $\tilde{\Delta}_\nu^\text{TED}$ using Viète's relations, which relate the roots of Eq.~\ref{eq:cubic} to its coefficients. At a TED, the relations, written in terms of $r$ for the repeated root and $s$ for the spectator root, become
\begin{align}
    2r + s &= 0, \\
    r^2 + 2rs &= \tilde{p}, \\
    r^2 s &= -\tilde{q}.
\end{align}
The residual splitting is $\tilde{\Delta}_\nu^\text{TED} = |r - s|$. Solving for $r$ and $s$ in terms of $\tilde{q}$ yields
\begin{equation}\label{eq:residual_splitting}
    \tilde{\Delta}_\nu^\text{TED} = \frac{3}{2^{1/3}} |\tilde{q}|^{1/3}.
\end{equation}
Expanding Eq.~\ref{eq:residual_splitting} at the TPD in the nuisance direction $\boldsymbol{\tilde{\delta}}$ gives the residual splitting as a function of nuisance perturbation:
\begin{align}
    \tilde{\Delta}_\nu^\text{TED}(\phi = 0) &= \frac{3 \cdot 2^{2/3} \left|\tilde{\kappa}_c^2 - 4\right|^{1/3}}{4} \left|\tilde{\delta}(\tilde{\Delta}_f)\right|^{1/3}, \label{eq:min_nu_growth_phi_0} \\
    \tilde{\Delta}_\nu^\text{TED}(\phi = \pi) &= \frac{3\tilde{\kappa}_c^{1/3} \left(\tilde{\kappa}_c^2 + 4\right)^{1/6}}{2} \left|\tilde{\delta}(\tilde{\Delta}_\kappa)\right|^{1/3}. \label{eq:min_nu_growth_phi_pi}
\end{align}
Equations \ref{eq:min_nu_growth_phi_0}-\ref{eq:min_nu_growth_phi_pi} exhibit the same cube-root nuisance dependence as Eqs. \ref{eq:mu_growth_phi_0}-\ref{eq:mu_growth_phi_pi}. $\tilde \Delta_\nu^\text{TED}$ is depicted in Fig. \ref{supp_fig:tne} (dashed green line). 

\subsection{\label{appendix-subsec:TED_str}Square-Root Splitting at TEDs}

Here we show that sensing paths crossing a TED retain square-root splitting. For fixed $\tilde{\kappa}_c$ and $\phi$, the discriminant is a scalar field $\mathrm{Disc}(\boldsymbol{\tilde{\Delta}})$ with $\mathrm{Disc}(\boldsymbol{\tilde{\Delta}}^\text{TED}) = 0$. For a sensing perturbation $\boldsymbol{\tilde{\epsilon}}$ crossing the TED, the discriminant can be Taylor expanded to first order as
\begin{equation}\label{eq:disc_sin}
    \mathrm{Disc}(\boldsymbol{\tilde{\Delta}}^\text{TED} + \boldsymbol{\tilde{\epsilon}}) \approx \lVert\nabla_{\boldsymbol{\tilde{\Delta}}}\mathrm{Disc}^\text{TED}\rVert \sin(\alpha) \lVert\boldsymbol{\tilde{\epsilon}}\rVert,
\end{equation}
where $\lVert\nabla_{\boldsymbol{\tilde{\Delta}}}\mathrm{Disc}^\text{TED}\rVert$ is the gradient over $\boldsymbol{\tilde{\Delta}}$ of $\mathrm{Disc}(\boldsymbol{\tilde \Delta^\text{TED}})$, and $\alpha$ is defined as the angle of incidence between the sensing path $\boldsymbol{\tilde \epsilon}$ and the tangent line to the local $\mathrm{Disc} = 0$ contour at the TED. Equation \ref{eq:disc_sin} is an approximate expansion, valid for $\lVert\boldsymbol{\tilde{\epsilon}}\rVert \ll \lVert\nabla_{\boldsymbol{\tilde{\Delta}}}\mathrm{Disc}^\text{TED}\rVert$, invalidated at the TPD where $\lVert\nabla_{\boldsymbol{\tilde{\Delta}}}\mathrm{Disc}^\text{TED}\rVert \to 0$.

Using the product form of the discriminant (Eq.~\ref{eq:disc_algebraic}), where $\mathrm{Disc} \propto \tilde{\Delta}_\nu^2$, and applying Viète's relations at the TED, we obtain
\begin{equation}\label{eq:ted_str}
\tilde{\Delta}_\nu(\boldsymbol{\tilde{\Delta}}^\text{TED} + \boldsymbol{\tilde{\epsilon}}) \approx \tilde{\Delta}_\nu^\text{TED} + \frac{\sqrt{\lVert\nabla_{\boldsymbol{\tilde{\Delta}}}\mathrm{Disc}^\text{TED}\rVert}}{3|\tilde{p}^\text{TED}|} \sqrt{\sin(\alpha)} \sqrt{\lVert\boldsymbol{\tilde{\epsilon}}\rVert},
\end{equation}
where $\tilde p^\text{TED} \neq 0$ and $\nabla_{\boldsymbol{\tilde\Delta}}\mathrm{Disc}^\text{TED} \neq 0$. The pre-factor to the square root term, identified as $\tilde a_\text{sqrt}^\text{TED}$, is finite for any non-grazing crossing ($\alpha \neq 0$), demonstrating that TEDs generally retain square-root splitting. Perpendicular crossings ($\alpha = \pi/2$) maximize the splitting strength. As $\alpha \to 0$, the square-root term vanishes and is replaced with either a constant, linear, or higher sublinear (such as cube-root) terms. The profile of $\tilde \Delta_\nu(\boldsymbol{\tilde{\Delta}}^\text{TED} +\boldsymbol{\tilde \epsilon})$ is depicted in Fig. \ref{fig:TED_single_panel}. 

\subsection{\label{appendix-sec:cusp_cat}Cusp Geometry and TED Location}

The nuisance scaling coefficient (Eq.~\ref{eq:nu_root_phi_0_nuisance}) is symmetric about the robust TPD, with $\tilde{\kappa}_c = 1.5$ and $\tilde{\kappa}_c = 2.5$ yielding identical values of $\tilde b_\text{cbrt}^\text{TPD}$. However, Fig.~\ref{fig:parametric} shows a clear asymmetry in $\sigma(\tilde{\nu}_\pm)$ between these configurations. This difference arises from the local cusp geometry. Specifically, nuisance perturbations shift not only the single peak frequency but also displace the TED location along the $\mathrm{Disc} = 0$ contour, leading to rapid change in $\lVert \boldsymbol{\tilde \Delta}^\text{TED} - \boldsymbol{\tilde \Delta}^\text{TPD} \rVert$. For $\phi = 0$ and $\phi = \pi$, this scales as
\begin{align}
    \lvert\tilde{\Delta}_\kappa^\text{TED} - \tilde{\Delta}_\kappa^\text{TPD}\rvert &\approx \frac{3 \cdot 2^{1/3} \lvert\tilde{\kappa}_c^2 - 4\rvert^{2/3}}{4\sqrt{8 - \tilde{\kappa}_c^2}} \left|\tilde{\delta}(\tilde{\Delta}_f)\right|^{2/3}, \label{eq:mu_growth_phi_0} \\
    \lvert\tilde{\Delta}_f^\text{TED} - \tilde{\Delta}_f^\text{TPD}\rvert &\approx \frac{3\tilde{\kappa}_c^{1/3}}{2(\tilde{\kappa}_c^2 + 4)^{1/6}} \left|\tilde{\delta}(\tilde{\Delta}_\kappa)\right|^{2/3}. \label{eq:mu_growth_phi_pi}
\end{align}
Equation \ref{eq:mu_growth_phi_0} is valid for $\phi = 0$, Eq. \ref{eq:mu_growth_phi_pi} for $\phi = \pi$. Notably, Eq.~\ref{eq:mu_growth_phi_0} vanishes at $\tilde{\kappa}_c = 2$, the robust TPD configuration. 

\section{\label{appendix:d} Appendix D: Experimental Methods}

\begin{figure*}[ht!]
\begin{centering}
\includegraphics[width=0.98\textwidth]{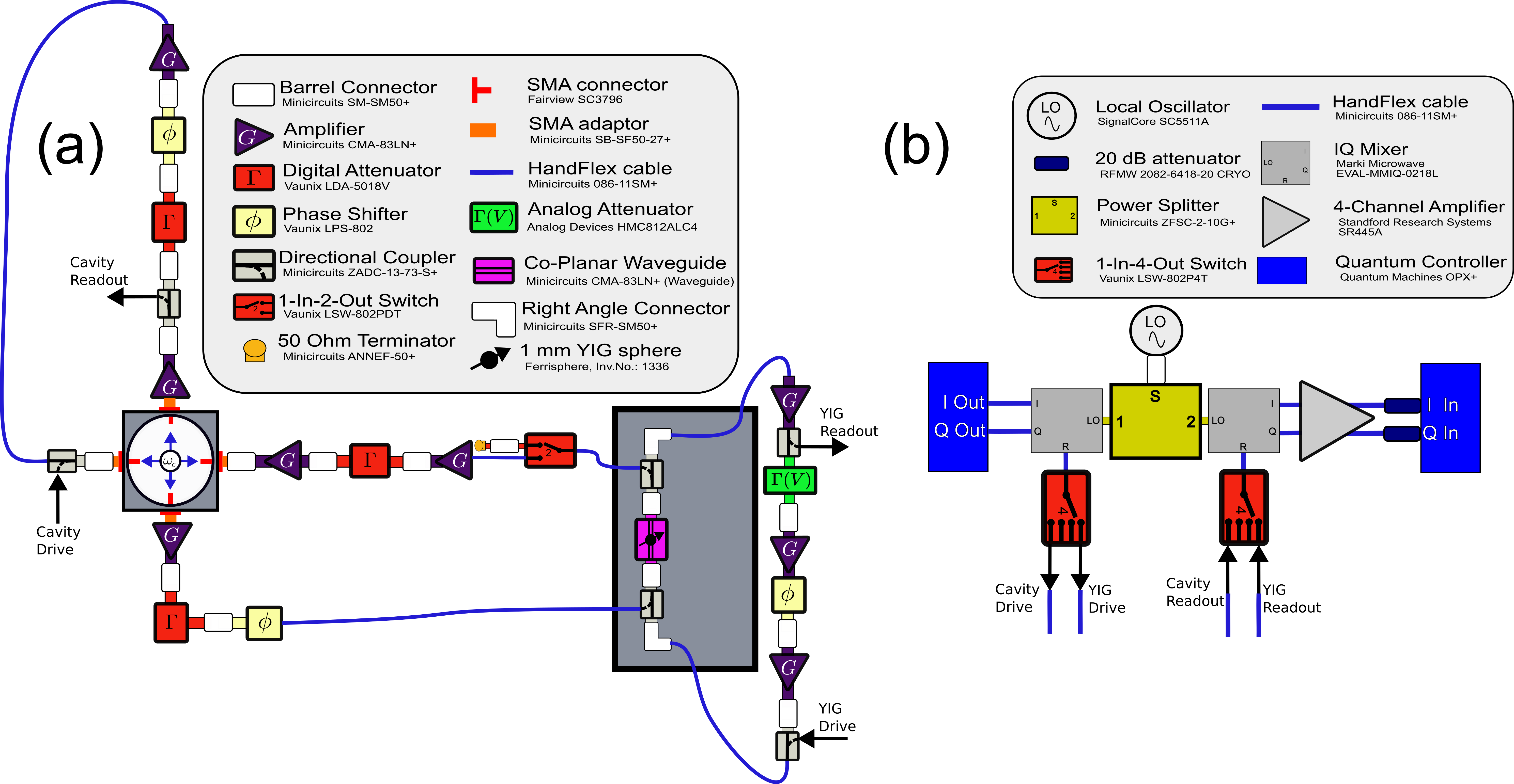}
\end{centering}
\caption{Experimental implementation. (a) Closed-loop configuration of coupled cavity and YIG mode, including switches to change between drive/readout operating modes, self-feedback loops, and all experimental devices. All devices, except for the amplifier, are controlled using a Python API. (b) Schematic of the microwave drive and readout circuitry, utilizing IQ mixers and a shared local oscillator to enable signal generation and detection.}
\label{supp_fig:expr_diagram}
\end{figure*}

\subsection{\label{appendix-sec:methods-device} Device Description}

Our measurement setup includes a vector network analyzer (Quantum Machines OPX+ Quantum Controller) and a local oscillator (SignalCore SC5511A) for collecting the data displayed in the Main Text. Our apparatus consists of a custom-built microwave cavity with mechanically tunable frequency and output-port coupling rates and a custom-built oscillator realized from a 1 mm diameter YIG sphere, placed on top of a co-planar waveguide (Minicircuits CMA-83LN+ with evaluation board TB-994+). The YIG sphere is coupled to the ground using orthogonal wire bonds. Magnon and photon modes are coupled over a meter-scale distance using flexible SMA wires (Minicircuits 086-11SM+). To achieve variable gain amplification and tunable coupling, we place a fixed-gain amplifier (Minicircuits CMA-83LN+), directly followed by a digital attenuator (Vaunix LDA-5018V), and tune the attenuation digitally using the Vaunix Python API. A digital phase shifter (Vaunix LPS-802) modifies the phase for hopping between the cavity and YIG. 

We use three digital switches to probe each mode individually during experimental operation without modifying the apparatus. Two Vaunix LSW-802P4T switches (1-input, 4-output) are used on the drive and readout lines to select which mode is being driven or read out. A 1-input, 2-output (Vaunix LSW-802PDT) switch is included on one of the coupling wires to disable or enable coupling when probing each mode individually. When disabled, the signal is routed directly to a 50-$\Omega$ terminator to ensure impedance matching and reduce reflections. Furthermore, when coupling is disabled, all digital attenuators are set to their maximum attenuation (50 dB) to provide additional isolation. Directional couplers (Minicircuits ZADC-13-73-S+) drive and read out each mode, attaching to the self-feedback loops. 

Experimental control of parameters is achieved through a Python API. $f_y$ is primarily tuned using an external magnetic field controlled by an electromagnet driven by a digital current source (SRS CS580). We offset the frequency of the YIG to 6 GHz using a fixed neodymium magnet and finely tune $B_z$ using the current source. To modify $\kappa_y$ with high resolution, we insert an analog voltage-controlled attenuator (Analog Devices HMC812ALC4 integrated with EV1 eval board) on the YIG self-feedback loop, and modify the voltage using precision voltage source (SRS DC205). A full schematic of the experimental apparatus is depicted in Fig.~\ref{supp_fig:expr_diagram}.

\subsection{\label{appendix-sec:methods-calibration} Parameter Calibration}

During an experimental sweep, all the parameters $f_c, f_y, \kappa_c, \kappa_y, J, \phi$ are necessary for data analysis. Parameters $f_c, f_y, \kappa_c, \kappa_y$ are measured during each step of the independent variable, using the RF-switch-based probing method described in Sec. \ref{appendix-sec:methods-device}. To determine $\phi$, we use the experimental apparatus to set $f_c = f_y$ and $\kappa_c = \kappa_y$, then couple the two modes together and change the digital phase shifter until the hybridized peaks are the same height in transmission, which defines the phase origin, $\phi = 0$. Finally, $J$ is a fitted parameter, extracted once for each value of loop attenuation and loop phase.

\subsection{\label{appendix-sec:methods-control} Parameter Control System}

To follow arbitrary paths in $\boldsymbol{\Delta}$-space, we model the relationship between coil current ($I$), analog attenuator voltage ($V$), and the measured mode parameters $(f_y, \kappa_y)$ using a simple system identification method. The pair $(I, V)$ are swept over a dense grid of $N = 1200$ points. At each $(I, V)$ we measure $(f_y, \kappa_y)$ via a Lorentzian fit of transmission. To capture the weak nonlinearities of the voltage-controlled attenuator we use the feature vector $\vec{\psi}(I, V) = [I, V, IV, V^2, 1, V]^T$ and perform least-squares regression using NumPy to obtain a weight matrix $\Theta \in \mathbb{R}^{6\times 2}$ which maps features to predicted outputs as
\begin{equation}
\begin{bmatrix}
\hat{f}_y \\
\hat{\kappa}_y
\end{bmatrix}  = \Theta^T \cdot \vec{\psi}(I,V).
\end{equation}
The model achieves $R^2 \approx 0.995$ for both $I$ and $V$ and RMSE on the order of $\approx 10 \text{ kHz}$. 

During an experimental sweep, this model serves as a feedforward map from control inputs $(I,V)$ to predicted outputs $(\hat{f}_y, \hat{\kappa}_y)$. The inverse problem is solved via Newton’s method to find the input required to attain a specified target.  At each iteration, a new measurement of $(f_y, \kappa_y)$ is taken, and every 5 iterations, a new measurement of $(f_c, \kappa_c)$ is taken to account for noise in the cavity. We calculate $(\Delta_f, \Delta_\kappa)$ and the absolute error relative to the target $(\Delta_f^\text{Target}, \Delta_\kappa^\text{Target})$ is computed, and a Jacobian is constructed analytically from the regression model. A damped Newton update with gain matrix $K_p = \mathrm{diag}([0.8, 0.8])$ is applied to $(I,V)$, and the process repeats until the error is within a specified tolerance, $\Delta_f^\text{err}, \Delta_\kappa^\text{err} = [10^{-5}, 10^{-5}]$. This protocol enables tracking of arbitrary paths in $(\Delta_f, \Delta_\kappa)$ using only a learned model and analytical Jacobian. 

\subsection{\label{appendix-sec:methods-data-analysis} Data Analysis}

In the Main Text, we strictly use nondimensionalized parameters to present the data. However, when performing data analysis, we use the dimensional parameters, and scale by $J$ at the end. When probing the YIG or cavity individually using the RF-switch-based isolation method, we convert the output data extracted from the OPX+ Quantum Controller into a linear scale and fit the data to a Lorentzian profile. From the fit, we extract $f_c, f_y, \kappa_c, \kappa_y$, and $\sigma(f_c), \sigma(f_y), \sigma(\kappa_c), \sigma(\kappa_y)$ using the elements of the least squares regression covariance matrix. Then, for data where the coupling is enabled, we fit $\nu_\pm$ to the data using $f_c, f_y, \kappa_c, \kappa_y$, and $J$, reported as experimental data in the Main Text as $\tilde \nu_\pm$. To quantify the fit uncertainty of $\nu_\pm$, we fit the data $N = 10^4$ times drawing from a multivariate normal distribution in $f_c, f_y, \kappa_c, \kappa_y$ with uncertainties $\sigma(f_c), \sigma(f_y), \sigma(\kappa_c), \sigma(\kappa_y)$. For each shot in $N$, we calculate $\nu_\pm$, yielding a distribution $\mathcal{D}(\nu_\pm)$ with standard deviation $\sigma(\nu_\pm)$. The error bars in the Main Text are $\sigma(\nu_\pm)/J$, a measure of the propagation of uncertainty of the fit parameters to the peak locations of the coupled system. 

We also seek to obtain a measure of global uncertainty across the entire experiment, accounting for variations in $\kappa_c$ across an entire sweep. The average $\kappa_c$ is reported in the form $\tilde \kappa_c$, with experimental uncertainty $\sigma_\text{exp}(\tilde \kappa_c)$ equal to the standard deviation of $\kappa_c$ across the sweep. The location of EPs, TPDs, and instability transitions are calculated using $J$ and $\kappa_c$. $\sigma_\text{exp}(\kappa_c)$ propagates to the location of EPs, TPDs, and instability, which we report in the Main Text. For $\phi = 0$ and $\phi = \pi$, a closed form solution for the propagation of uncertainty from $\tilde \kappa_c$ exists. For other values of $\phi$, no reasonable closed form solution exists because the location of TPDs is defined by a quartic polynomial (See Appendix \ref{appendix:b}). In this case, a Monte Carlo propagation of uncertainty technique is used instead. Beyond the instability transition, the linear model is no longer valid, and the experimental results are no longer well-described by $\tilde \beta_{\text{ss}}$. An exploration of nonlinear phenomena in a similar system can be found in Ref. \cite{salcedo-gallo_demonstration_2025}.

\subsection{\label{appendix-sec:yig_ferro}YIG Ferromagnetic Resonance}

Although YIG is technically ferrimagnetic, we ensure the material is driven into magnetic saturation. In this regime, the net magnetization precesses uniformly, and the standard ferromagnetic formalism can accurately describe its dynamics. Thus, the Larmor frequency depends linearly on the applied field as $f_y = \mu_B B_z/2\pi$, where $\mu_B$ is the Bohr magneton, consistent with the simplified Kittel formula for a uniformly magnetized sphere \cite{sparks_ferromagnetic_1961}. Several factors ensure the validity of this approximation. First, the saturation magnetization of our YIG sample is $\mu_0M_s = 0.178$ T, and we apply a bias field exceeding this value using permanent neodymium magnets. Second, the YIG sample is spherical, eliminating shape-dependent demagnetization effects since the demagnetization factors are isotropic \cite{osborn_demagnetizing_1945}. Third, thermal effects can be neglected in our regime, as the Curie temperature of YIG ($\sim560 ^{\circ}$K) \cite{zhang_cavity_2015} is far above our room temperature operating conditions.

\section{\label{appendix:e} Data and Code Availability}

The data that support the findings of this study are available online on \href{https://github.com/Alex-Carney/EP_TPD_Unification}{GitHub}. Please follow instructions for downloading database files described in the repository description. Additionally, an online web application built using \href{https://ep-tpd-educational-app.streamlit.app/}{Streamlit} is available to explore the theory of EPs and TPDs. We thank Sam Sacerdote for his work on this application. The data analysis, figure generation code, and computer algebra derivations are available online on \href{https://github.com/Alex-Carney/EP_TPD_Unification}{GitHub}.

\end{document}